\documentclass[superscriptaddress,aps,pra,twocolumn,showpacs,nofootinbib,longbibliography,notitlepage]{revtex4-1}
\usepackage{etex}
\usepackage{amsmath,amssymb,amsthm}
\usepackage[colorlinks=true,citecolor=blue,urlcolor=blue]{hyperref}
\usepackage{times,txfonts}
\usepackage{braket}
\usepackage{color}
\usepackage{natbib}
\usepackage{amsmath,blkarray}
\usepackage{mathtools}
\usepackage{ulem}
\usepackage{latexsym}
\usepackage{tabularx, booktabs}

\usepackage{calc,graphicx,color,bm}
\usepackage{subfigure}
\usepackage{graphicx}

\usepackage{amsfonts}
\usepackage{tikz}

\newcommand{\beq}{\begin{equation}}
\newcommand{\eeq}{\end{equation}}
\newcommand{\beqa}{\begin{eqnarray}}
\newcommand{\eeqa}{\end{eqnarray}}

\usepackage{amsmath}
\usepackage{graphicx}
\usepackage{amssymb}
\usepackage{txfonts,color}
\makeatother

\begin{document}

\title{Time-optimal variational control of bright matter-wave soliton }

\date{\today}

\author{Tang-You Huang}
\affiliation{International Center of Quantum Artificial Intelligence for Science and Technology (QuArtist) \\ and Department of Physics, Shanghai University, 200444 Shanghai, China}
\affiliation{Department of Physical Chemistry, University of the Basque Country UPV/EHU, Apartado 644, 48080 Bilbao, Spain}

\author{Jia Zhang}
\affiliation{International Center of Quantum Artificial Intelligence for Science and Technology (QuArtist) \\ and Department of Physics, Shanghai University, 200444 Shanghai, China}


\author{Jing Li}
\affiliation{Quantum Systems Unit, Okinawa Institute of Science and Technology Graduate University, Onna, Okinawa 904-0495, Japan}

\author{Xi Chen}
\email{xchen@shu.edu.cn} 
\affiliation{International Center of Quantum Artificial Intelligence for Science and Technology (QuArtist) \\ and Department of Physics, Shanghai University, 200444 Shanghai, China}
\affiliation{Department of Physical Chemistry, University of the Basque Country UPV/EHU, Apartado 644, 48080 Bilbao, Spain}

\begin{abstract}
	
Motivated by recent experiments, we present the time-optimal variational control of bright matter-wave soliton trapped in a quasi-one-dimensional harmonic trap by manipulating the atomic attraction through Feshbach resonances. More specially, we first apply a time-dependent variational method to derive the motion equation for capturing the soliton's shape, and secondly combine inverse engineering with optimal control theory to design the atomic interaction for implementing time-optimal decompression. Since the time-optimal solution is of bang-bang type, the smooth regularization is further adopted to smooth the on-off controller out, thus avoiding the heating and atom loss, induced from magnetic field ramp across a Feshbach resonance in practice. 

\end{abstract}
\maketitle

\section{Introduction}

The experimental discovery of Bose-Einstein condenstates (BECs) in 1995 has instigated a broad interest in ultracold atoms and molecules \cite{anderson1995observation,bradley1995evidence,davis1995bose}, and paved the way for extensive studies on the nonlinear properties and dynamics of Bose gases, with the applications in atom optics and other areas of condensed
matter physics and fluid dynamics \cite{dalfovo1999theory}. For atomic matter waves,
the matter-wave soliton can be experimentally created in BECs with repulsive and attractive interaction between atoms which indicates dark soliton \cite{DarkScience2001,DarkPRL.83.5198} and bright soliton \cite{khaykovich2002formation,strecker2002formation} respectively. Subsequently, more experimental findings show the formation of bright solitary matter-waves and probe for potential barriers \cite{marchant2013controlled,PhysRevA.93.021604}. 
 Very recently, the bright solitons
are created by double-quench protocol, that is, by
a quench of the interactions and the longitudinal confinement \cite{Excitation2019prl}.
In this regard, bright solitons, i.e. nonspreading localized wave packet,
are the most striking paradigm of nonlinear
system, since bright soliton and bright solitary waves are 
the excellent candidates for the applications in highly sensitive atom interferometry \cite{martin2012quantum,Sagnac,Interferometerexp} or the generation of Bell state in quantum information processing \cite{Bellstate}. 

In the mean field approximation, an atomic BEC obeys the  Gross-Pitaevskii (GP) equation, which is equivalent to the three-dimensional (3D) nonlinear Schr\"{o}dinger equation.
While in quasi-one-dimensional (1D)  regime,
these systems with BECs confined in a cigar-shaped potential trap are reduced to the 1D GP equation \cite{3Danisotropic}. 
In particular, with the experimental feasibility of reaching the quasi-1D limit of true solitons, 
the modulation of the scattering length by varying the magnetic field through a board Feshbach resonance, gives rise to prominent nonlinear features, such as collapse \cite{donley2001dynamics,PhysRevLett.96.170401},  collision \cite{nguyen2014collisions} and instability \cite{nguyen2017formation}. In most aforementioned experiments \cite{khaykovich2002formation,strecker2002formation,marchant2013controlled,PhysRevA.93.021604,PhysRevLett.96.170401,Excitation2019prl,nguyen2017formation}, the quenching of atom interactions
from repulsive to attractive makes the cloud
unstable, resulting in the excitation of breathing modes \cite{breathingmode}. Meanwhile, the experimentally observed atom loss rate, relevant to inelastic three-body collisions, becomes the orders of magnitude larger than one would expect for static soliton \cite{collective}. 
Therefore, shortcuts to adiabaticity (STA) \cite{torrontegui2013shortcuts,STARMP} is requested to
surpass the common non-adiabatic process, for instance, thus avoiding the significant heating and losses, induced from the sudden switching of the atomic interactions \cite{Thomas}.

By now, variational technique, originally proposed in nonlinear problem \cite{variationalprl,variationalpra}, have been developed for STA in particular systems \cite{li2016shortcut,jingnpj,xu2019quantum,tangyou} that cannot be treated by means of other existing approaches, i.e. invariant-based engineering \cite{muga2009frictionless,chenprl104}, counterdiabatic driving \cite{berry2009transitionless,adolfocddriving,adolfoprx}, and 
fast-forward scaling \cite{masuda2008fast,fastforward}.
More specifically, since the time-dependent variational principle can find a set of Newton-like ordinary differential equations for the parameters (i.e.  the width of cloud,  center and interatomic interaction), 
the variational control provides a promising alternative, aiming at accelerating the adiabatic compression/decompression of BECs and bright solitons \cite{li2016shortcut,tangyou}, beyond the harmonic approximation of the potential \cite{xu2019quantum} and Thomas-Fermi limit \cite{muga2009frictionless,StefanatosLiPRA2012,PhysRevR2020}. In this scenario, the Lewis-Riesenfeld dynamical invariant and general scaling transformations \cite{muga2009frictionless,chenprl104}
are not required in the context of inverse engineering.

In this article, we shall address the time-optimal variational control, by focusing on the bright matter-wave solitons with the tunable atomic interaction in harmonic trap \cite{LiuPRL,CastinPRA,Lucapra}. Here we first hybridize the variational approximate and inverse engineering methods to design the STA, and further apply 
the Pontryagain's Maximum principle in optimal control theory \cite{kirk2004optimal} for achieving the time-minimal decompression, fulfilling the appropriate boundary conditions. Under the constraint on atomic interaction, time-optimal solution delivers bang-bang control, which requires the dramatic changes in the interaction strength through rapid tuning of an external magnetic field around a Feshbach resonance. It turns out that such sudden change leads to the heating and atom loss, excites the breathing modes, and thus make the practical experiment unstable or unfeasible \cite{nguyen2017formation,collective}. Therefore, this motivates us to try the smooth regularization of bang-bang control at the expense of operation time \cite{yongchengPRA,silva2010smooth}. Our results are of interest to deliver a fast but stable creation or transformation of soliton \cite{collective,Excitation2019prl,nguyen2017formation}, 
and have the fundamental implications for quantum speed limit and thermodynamic limits of atomic cooling \cite{jingnpj,xu2019quantum,tangyou}. 

\section{Variational method of soliton dynamics}

We consider a BEC of $N$ atoms of mass $m$ and attractive
s-wave scattering length $a_s <0$, trapped in a prolate, cylindrically symmetric harmonic trap \cite{3Danisotropic,LiuPRL,CastinPRA,Lucapra}. To be consistent, we write down the dynamics of a BEC described by the following time-dependent 3D GP equation:
\beq
\label{3DGP}
\left[i \hbar \frac{\partial}{\partial t}+\frac{\hbar^2}{2m} \nabla^2 - U(r) -g_{3D}(t) |\Psi|^2 \right] \Psi=0, 
\eeq%
where $\Psi(r,t)$ is the macroscopic wave function (order parameter)
of BEC, $g_{3D}(t)=4 N \pi \hbar^2 a_s(t)/m$ is the   interactomic strength, proportional to controllable   s-wave scattering length $a_s(t)$, and the harmonic trap modeled by 
\beq
U(r)= \frac{1}{2}m [\omega^2 x^2+ \omega^2_{\perp}(y^2+z^2)],
\eeq
with the static longitudinal and transverse trapping frequencies being $\omega$ and $\omega_{\perp}$. Here the time-dependent $a_s(t)$ can be modulated by the external magnetic field through a Feshbach resonance for our proposal. 

For sufficiently tight radial confinement ($\omega \ll \omega_{\perp} $), it is reasonable to assume a reduction to a quasi-1D GPE equation by using the wave function \cite{Lucapra},
\beq
\label{3Dwavefunction}
\Psi(r,t)= \psi(x,t)\exp[-(y^2+z^2)/2\sigma_{\perp}]/\sqrt{\pi \sigma^2_{\perp}},
\eeq
with $\sigma_{\perp}= \sqrt{\hbar/m \omega_{\perp}}$ being the transverse
width, when the traverse energy $E_{\perp}= \hbar \omega_{\perp}$.
By substituting Eq. (\ref{3Dwavefunction}) into Eq. (\ref{3DGP}) and integrating the underlying 3D GP equation in the transverse directions, we obtain 
\beq
\label{1DGP}
\left[i \hbar \frac{\partial}{\partial t}+\frac{\hbar^2}{2m} \frac{\partial^2}{\partial x^2} -E_{\perp}- \frac{1}{2} m \omega^2 x^2 -g_{1D}(t) |\psi|^2 \right] \psi=0, 
\eeq%
with $g_{1D}(t)= g_{3D}(t)/2\pi \sigma^2_{\perp}$.  For convenience, we introduce 
the dimensionless variables with tildes in physical units:
$\tilde{t} = \omega_{\perp} t$,
$\tilde{\omega}=\omega/\omega_{\perp}$, $\tilde{x}= x/\sigma_{\perp}$,  $\tilde{g}(t)=g(t) /\hbar\omega_{\perp}\sigma_{\perp}$ with imposed $g(t) \equiv g_{1D}(t)= 2 N\hbar \omega_{\perp} a_{s}(t)$, such that the reduced 1D GPE equation for wave function $\psi(x,t)$ along the longitudinal direction reads
	%
\beq
\label{GP-equation}
i \frac{\partial \psi }{\partial t}=-\frac{1}{2}\frac{%
		\partial ^{2} \psi }{\partial x^{2}}+\frac{1}{2} \omega ^2 x^2 \psi + g(t) |\psi |^{2}\psi.
	\eeq%
Here all variables are dimensionaless, and we ignore the tilde notation from now on, for simplicity. 

Since the 1D nonlinear Schr\"{o}dinger equation supports
the ground state in the form of a bright soliton, we consider 
the standard sech ansatz, instead of Gaussian ansantz,
\begin{eqnarray}
\psi(x,t)=A(t)\mathrm{sech}\left[\frac{x}{a(t)}\right] e^{i b(t)x^2},
\label{sech}
\end{eqnarray}
for describing the dynamics, where the amplitude
$A(t)=\sqrt{N/2a(t)}$ is normalized by 
$\int^{+\infty}_{-\infty}|\psi|^{2} dx=2 a(t) A^{2}(t)=N$,
$a(t)$ is the longitudinal size of atomic size,
and $b(t)$ represents the chirp and have the relevance to 
currents. In order to apply the time-dependent variational principle \cite{variationalprl,variationalpra}, we write down
the Lagrangian density $\mathcal{L}$,
\begin{equation}	
\label{Lagrangian}
\mathcal{L}=\frac{i}{2}\left( \frac{\partial \psi }{\partial t}\psi ^{\ast }-\frac{%
	\partial \psi ^{\ast }}{\partial t}\psi \right) -\frac{1}{2}|\frac{\partial
		\psi }{\partial x}|^{2}-\frac{1}{2}g(t)|\psi |^{4}-\frac{1}{2} \omega x^2|\psi |^{2},~~
	\end{equation}
where the asterisk denotes complex conjugation.
Inserting Eq. (\ref{sech}) into Eq. (\ref{Lagrangian}), we calculate a grand Lagrangian by integrating the Lagrangian density over the whole coordinate space, 
$L=\int^{+\infty}_{-\infty} \mathcal{L}dx$. Applying the Euler-Lagrange formulas $\delta L/\delta p=0$, where $p$ presents one of the parameters $a(t)$ and $b(t)$, we obtain  $b=\dot{a}/2a(t)$ and the following differential equations:
\beq
\label{Ermakov}
\ddot{a}+\omega^2 a(t)=\frac{4}{\pi^2 a^3(t) }+\frac{2g(t)}{\pi^2a^2(t)}.
\eeq
This resembles the generalized Ermakov equation \cite{chenprl104,tangyou}, which can be exploited 
to design STA based on the inverse engineering with the appropriate boundary conditions. The main difference from previous results is that we concentrate on the the time modulation of atomic interaction, instead of trap frequency.
In what follows we shall concern about the design STA 
by quenching the atomic interaction, within minimal time. 

	
\section{Shortcuts to Adiabaticity}
	
The generalized Ermakov equation (\ref{Ermakov}) is analogous to Newton's second differential equation for a fictitious particle with unit mass, with effective potential,
\beq\label{U(a)}
U(a) = \frac{1}{2}\omega^2a^2+\frac{2}{\pi^2a^2}+\frac{2g(t)}{\pi^2 a},
\eeq
as found in Landau’s mechanics \cite{landau1998course}. In general, the dynamic equation for the width $a(t)$ provides the analytical treatment of collective mode when ramping the atom-atom interaction suddenly, $g(t) \to 0$ \cite{collective}. Here we aim to apply inverse engineering to design the interaction
for realizing the speed up of adiabatic expansion, when the experimental resolution is improved by creating a bright soliton with a larger longitudinal width \cite{khaykovich2002formation,Lucapra}. Of course, the result can be directly extended to soliton compression \cite{abdullaev2003adiabatic,li2016shortcut} without any efforts.

In this vein, we consider the fast transformation from the initial state at $t = 0$ to the target one at $t = \tau$, keeping the shape invariant, where the initial width $a(0) =  a_{\mathrm{i}}$ ends up with 
the targets $a(\tau) = a_{\mathrm{f}}$ by adjusting the interaction from $g(0) = g_{\mathrm{i}} $ 
to $g(\tau) = g_{\mathrm{f}}$. Here $ a_{\mathrm{f}} > a_{\mathrm{i}}$ ($a_{\mathrm{f}} < a_{\mathrm{i}}$) implies the decompression (compression). 
 To this end, we first introduce the
the boundary conditions,
\beqa
\label{Bounday condition-1}
a(0) &=&  a_{\mathrm{i}}, ~ a(\tau) =  a_{\mathrm{f}},
\\
\label{Bounday condition-2}
\dot{a}(0) &=& \dot{a}(\tau)=0,
\\
\label{Bounday condition-3}
\ddot{a}(0) &=& \ddot{a}(\tau)=0,
\eeqa
where $a_{\mathrm{i}}$ and $a_{\mathrm{f}}$ are determined by the following equation 
\beqa
\label{adiabaticref}
a^4 -\frac{2  g(t) }{\pi^2 \omega^2} a=\frac{4}{\pi^2 \omega^2},~
\eeqa
when $g(t)$ is specified by initial and final values, $g(0) = g_{\mathrm{i}} $ and $g(\tau) = g_{\mathrm{f}}$. Eq. (\ref{adiabaticref}) is so-called adiabatic reference, 
resulting from Eq. (\ref{Ermakov}) when the condition $\partial U/\partial a=0$, yielding $\ddot{a}=0$, is considered. This is analogous to perturbative Kepler problem~\cite{landau1998course}, which actually indicates the fictitious particle stays adiabatically at the minimum of effective potential $(\ref{U(a)})$ . 

With boundary conditions (\ref{Bounday condition-1}-\ref{Bounday condition-3}), we apply the inverse engineering based on Eq. (\ref{Ermakov}). 
In order to exemplify STA, we choose a simple polynomial ansatz,
\beq
a\left(t\right)= a_{\mathrm{i}}-6( a_{\mathrm{i}}- a_{\mathrm{f}})s^{5}+15( a_{\mathrm{i}}- a_{\mathrm{f}})s^{4}-10( a_{\mathrm{i}}- a_{\mathrm{f}})s^3,
\eeq
with $s = t/\tau$ and $\tau$ being the total time, fulfilling the all boundary conditions. After we interpolate the function of $a(t)$, the interaction $g(t)$ is eventually designed 
from Eq. (\ref{Ermakov}). The designed interaction $g(t)$ is smooth, and the switching of the scattering length 
can be easily implemented in the experiments \cite{nguyen2017formation,khaykovich2002formation}. In principle, the total time $\tau$ can be arbitrarily short from the viewpoint of mathematics. The polynomial ansatz is simple but not optimal at all. We are planning to address 
the time-optimal control problem with the physical constraint on the interatomic interaction .

\section{Time-Optimal control and Smooth Regularization}

\subsection{``bang-bang” control}
Next, we formulate the minimum time control according to the
Pontryagain's Maximum principle in optimal control theory \cite{kirk2004optimal}. For brevity, we introduce $x_1(t) = a$, $x_2(t) = \dot{a}$, and rewrite the dynamics of system from (\ref{Ermakov}) into two first-order differential equations:
\beqa
\dot{x}_1&=&x_2, \label{differential-x1} \\
\dot{x}_2&=&-\omega^2x_1+\frac{4}{\pi^2x_1^3}+\frac{2u(t)}{\pi^2x_1^2}, \label{differential-x2}
\eeqa
where the bounded control function $u(t) = g(t)$.
Without loss of generality, we may simple choose  $a_{\mathrm{i}} =1$, $a_{\mathrm{f}}= \gamma$, $g_{\mathrm{i}}<0$ and $g_{\mathrm{i}}<g_{\mathrm{f}}$, when $\gamma>1$ is considered for the decompression
of bright soliton with tunable interaction. In this context,
we formulate the time-optimal problem that drives
the state $\textbf{x}_i =\{x_1(t),x_2(t)\}$ from the initial 
$\{1,0\}$ to final $\{\gamma,0\}$, 
under the constrain $g_\mathrm{i} \leq u(t) \leq g_\mathrm{f}$.
  
\begin{figure}[]
	\includegraphics[width=\columnwidth]{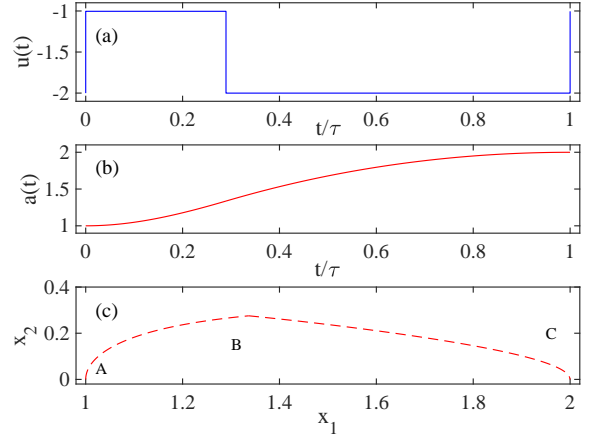}
	\caption{(a) Controller $u(t)$ of ``bang-bang " type, for the time-optimal control of soliton decompression.  (b) The evolution of $a(t)$, the width of bright soliton, is depicted.  
		(c) The trajectory of $(x_1,x_2)$, where the initial point $A= (1,0)$, intermediate point $B=(x_1^B,x_2^B)$ and $C=(\gamma,0)$
		are illustrated.  Parameters are: $\omega = \tilde{\omega}/\omega_{\perp} = 0.01$
		(transverse trapping frequency $\omega_{\perp} =250 \times 2\pi$ Hz), $\gamma = 2$, $g_{\mathrm{i}} =-2.0005$, $g_{\mathrm{f}}=-1.0039$, and $\tau= 7.0183$ with the switching time $t_1=2.0325$.}
	\label{fig1}
\end{figure}

To find the minimal time $\tau$, we define the cost function,
\beq
J \equiv \int_{0}^{\tau} dt = \tau.
\eeq%
The control Hamiltonian $H_c (\textbf{p}, \textbf{x}, u) $ is defined as:
\beq
\label{controlH}
H_{c} (\textbf{p}, \textbf{x}, u) = p_0 +p_1 x_2-p_2\omega^2x_1 
+\frac{4p_2}{\pi^2x_1^3} 
+\frac{2p_2u(t)}{\pi^2x_1^2}
\eeq%
where $\textbf{p}_i= (p_0, p_1, p_2)$ are non-zero and continuous Lagrange multipliers, $p_0 < 0$ can be chosen for convenience since it amounts to multiplying the cost function by a constant, and $\textbf{p}_i$ fulfill the Hamilton's equations, $\dot{\textbf{x}}=\partial H_{c}/\partial \textbf{x}$ and $\dot{\textbf{p}}=-\partial H_{c}/\partial \textbf{x}$. For almost all $0 \leq t \leq \tau$, the function $H_c(\textbf{p},\textbf{x},u)$ attains its maximum at $u= u(t)$, and $H_c(\textbf{p},\textbf{x},u)=c$, where $c$ is constant. With the help of the Hamiltonian's equation, we have explicit expression,
	\beqa\label{pi}
	\dot{p}_1 &=& p_2 (\omega^2 + \frac{12}{\pi^2x^4_1}+\frac{4u}{\pi^2x_1^3}),
	\\
	\dot{p}_2 &=& -p_1.
	\eeqa 
It is clear that the control Hamiltonian $H_c(\textbf{p},\textbf{x},u)$ is a linear function of the control variable $u(t)$. Therefore, the maximization of $H_c(\textbf{p},\textbf{x},u)$ is determined by the sign of the term $2p_2u(t)/\pi^2x_1^2$, which is only related with $p_2$, since the width, $a(t)$, is always positive, i.e. $x_1>0$, and $p_2\not= 0$. Here $p_2=0$ does not provide the singular control, and only happens at specific instant moments (switching times) \cite{LuPRA2014}, and we set $\delta = g_\mathrm{f}$. 
Thus, we can obtain $u(t) =g_{\mathrm{f}}$ when $p_2>0$ at time $t\in (0,t_{1})$, and $u(t) = g_{\mathrm{i}}$ when $p_2<0$ at time $t\in ( t_{1}, t_1+t_2)$, such that the controller has the 
form of  ``bang-bang" type, see Fig. \ref{fig1}(a),
\beq
\label{control sequence u(t)}
u(t)=%
\begin{cases}
	g_{\mathrm{i}}, \qquad  t = 0\\
	~g_{\mathrm{f}}, \qquad  0<t<t_1 \\
	~g_{\mathrm{i}}, \qquad  t_1\leq t<t_1+t_2\\
	g_{\mathrm{f}},  \qquad  t = t_1 + t_2 = \tau
\end{cases}.
\eeq%
As a consequence, the time-optimal control suggests 
the abrupt changes of controller at the switching times. 
When control function $u$ is constant, from Eqs. (\ref{differential-x1}) and (\ref{differential-x2}), one can find $x_1$ and $x_2$ satisfies
\beq\label{x1-x2}
x_2^2 + \omega^2x_1^2+\frac{4}{\pi^2x_1^2}+\frac{4u}{\pi^2x_1}=c,
\eeq
with constant $c$. With the ``bang-bang" protocol of controller (\ref{control sequence u(t)}), the system evolves from the initial point $A(1,0)$, along intermediate one $B(x_{1}^{B},x_{2}^{B})$, and finally end up with the target point $C(\gamma,0)$, in the phase space $(x_1,x_2)$.

Now we manage to calculate the times for two segments, $AB$
and $BC$, by substituting  $u =g_{\mathrm{f}}$ or $u =g_{\mathrm{i}}$ into dynamical equations (\ref{differential-x1}) and (\ref{differential-x2}), respectively.
Thus, we have the equation for the first segment $AB$ for $t\in ( 0,t_{1})$,
\beq\label{first-seg}
x_2^2+\omega^2x_1^2+\frac{4}{\pi^2x_1^2}+\frac{4g_{\mathrm{f}} }{\pi^2x_1}=c_1,
\eeq
with $c_1 = \omega^2+4/\pi^2+4g_{\mathrm{f}} /\pi^2$, and the second segment $BC$ for $t\in \lbrack t_{1},t_{1}+t_{2})$
\beq\label{second-seg}
x_2^2+\omega^2x_1^2+\frac{4}{\pi^2x_1^2}+\frac{4g_{\mathrm{i}} }{\pi^2x_1}=c_2,
\eeq
with $c_{2} = \omega^2\gamma^2+ 4/\pi^2\gamma^2+4g_{\mathrm{i}} /\pi^2\gamma$.  The matching condition for the intermediate point $B(x_{1}^{B},x_{2}^{B})$ yields 
\beq\label{x1B}
x_1^B=\frac{8\delta ^2\gamma^2}{\left(\gamma+1\right)\left[\left(\gamma-1\right)\left(4-\omega^2\pi^2\gamma^2\right)+4\delta\gamma\right]},
\eeq
from which we can determine the switching time $t=t_1$ and the total time $\tau=t_1 + t_2$ as follows,
\beq
\label{tf}
\tau=t_{1}+t_{2}, 
\eeq
where  
\beqa
\label{t1}
t_{1} &=& \int_{\beta}^{x_B}\frac{dx}{\sqrt{c_1-\omega^2x^2-4/\pi^2x^2-4
	g_{\mathrm{f}}/\pi^2x}},
\\
\label{t2}
t_{2} &=& \int_{x_B}^{\gamma}\frac{dx}{\sqrt{c_2-\omega^2x^2-4/\pi^2x^2-4g_{\mathrm{i}} /\pi^2x}}.
\eeqa

\begin{figure}[t]
	\includegraphics[width=3.0in]{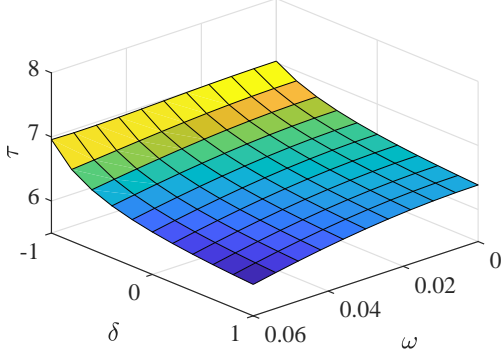}
	\caption{ Minimum time $\tau$ versus trap frequency $\omega$ 
		and physical constraint $\delta $ for bright soliton decompression, where the parameters are the same as those in Fig. \ref{fig2}. }
	\label{fig2}
\end{figure}
	
Figure \ref{fig1} illustrate the trajectory of $(x_1, x_2)$, corresponding to the evolution of width $a$, by using the time-optimal solution of soliton decompression with the controller $u(t)$ of ``bang-bang" type. Here we take the parameters: $\omega = \tilde{\omega}/\omega_{\perp} = 0.01$
(transverse trapping frequency $\omega_{\perp} =250 \times 2\pi$ Hz), $\gamma = 2$, $g_{\mathrm{i}} =-2.0005$, and $g_{\mathrm{f}} =-1.0039$.
In this case, the minimal time is obtained as $\tau= 7.0183$,
with the switching time $t_1=2.0325$. Noting that the minimal time is different from the cooling process in time-dependent harmonic trap \cite{StefanatosLiPRA2012,StefanatosPRA2010,tangyou}, where the attractive interaction slows down the cooling process, thus decreasing the cooling rate of thermodynamic cycle \cite{tangyou}.

Furthermore, we display the effect of trap frequency  $\omega$ and the physical constraint on the minimum time $\tau$ in Fig. \ref{fig2}, where the controller $u(t)$ is bounded by $g_{\mathrm{i}} \leq u(t) \leq \delta$ and other parameters are the same as those in Fig. \ref{fig1}. 
We visualize
that when the same physical constraint is set, the minimal time $\tau$ decreases when trap becomes tight, corresponding to the large trap frequency. Meanwhile, the minimal time $\tau$ is decreased, and even approaches zero, when large constraint $\delta$ is allowed. In pursuit of shorter time in decompression process, the positive region is expected for the constrain $\delta$. 
Here we emphasize that the minimal time, depending on the trap frequency and atom-atom interaction, have fundamental implications to efficiency and power in quantum heat engine with bright soliton as working medium \cite{jingnpj}. Of course, the STA compression/decompression can replace the
adiabatic branches in quantum refrigerator, clarifying the third law of thermodynamics as well \cite{hoffmann2011time}.

So far, we attain the minimum-time control of bright-soliton decompression with ``bang-bang" type, see Eq. (\ref{control sequence u(t)}). This Heaviside function suggests the abrupt changes of interatomic interaction. However, the sudden change of s-wave scattering length makes the soliton decompression unstable. When the operation time is much shorter, the interaction has been changed rapidly from negative and positive by modulating an external magnetic field. This could lead to significant atom loss and heating across a Feshbach resonance. 
 
 \begin{figure}[]
 	\includegraphics[width=\columnwidth]{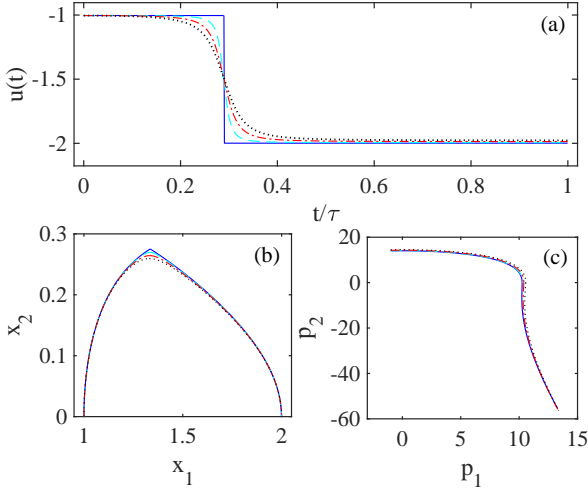}
 	\caption{ (a) Smooth controller $u^{\epsilon}(t)$ with different values, $\epsilon = 0$ (blue solid curve), $\epsilon = 0.1$ (cyan dashed curve), $\epsilon = 0.2$ (red dash-dotted curve), and $\epsilon = 0.3$ (black dotted curve).
 	(b) The trajectory of $(x_1, x_2)$,
 		where the initial point $A=(1,0)$, intermediate point $B=(x_1^B,x_2^B)$ and final point $C=(\gamma,0)$ are illustrated, with the related Lagrange multipliers $(p_1, p_2)$ in (c). The fixed $\delta^{\epsilon}$ is listed in Table. \ref{table1}, and other parameters are the same as those in Fig. \ref{fig2}.}
 	\label{fig3}
 \end{figure}
 
\subsection{smooth regularization}

Inspired by smooth regularization \cite{silva2010smooth}, we reformulate the control function $u(t)$ to $u^{\epsilon}(t)$ by introducing a real small constant $\epsilon$ to avoid the dramatic change in the controller. For this purpose, the system and controller are labeled by the superscript $\epsilon$, yielding the new continuous controller $u^{\epsilon}(t)$,
and the regularized control system $\textbf{x}^{\epsilon}_i= (x^{\epsilon}_1, x^{\epsilon}_2)$ in the form of
\beqa
u^{\epsilon}(t) = \frac{(g_{\mathrm{i}}^{\epsilon}-\delta) p_2^{\epsilon}}{2\sqrt{(p_2^{\epsilon}(t))^2+\epsilon^2(p_1^{\epsilon}(t))^2}},
\label{u_fix}
\eeqa
and 
\beqa
\dot{x_1}^{\epsilon}&=&x_2^{\epsilon} ,
\label{x1dotfixed} \\
\dot{x_2}^{\epsilon}&=&-\omega^2x_1^{\epsilon}+\frac{4}{\pi^2(x_1^{\epsilon})^3}+\frac{2u^{\epsilon}(t)}{\pi^2(x_1^{\epsilon})^2}.
\label{x2dotfixed} 
\eeqa
These grantee that $u^{\epsilon}(t)$ reduces to $u(t)$, when
$\epsilon = 0$, as seen in the control of ``bang-bang" type  (\ref{control sequence u(t)}). In this scenario, we can have the similar control Hamiltonian $H_c(\textbf{p}^{\epsilon},\textbf{x}^{\epsilon},u^{\epsilon})$
as Eq. (\ref{controlH}). As a result, the differential equation of the Lagrange multipliers, $\textbf{p}^{\epsilon}_i= (p^{\epsilon}_0, p^{\epsilon}_1, p^{\epsilon}_2)$, is obtained as 
\beqa
\label{fixed p1}
\dot{p^{\epsilon}_1} &=& p^{\epsilon}_2 (\omega^2 + \frac{12}{\pi^2(x^{\epsilon}_1)^4}+\frac{4u_1^{\epsilon}}{\pi^2(x_1^{\epsilon})^3}),
\\
\label{fixed p2}
\dot{p^{\epsilon}_2} &=& -p^{\epsilon}_1.
\eeqa 
Here $x_1^{\epsilon}$ and $x_2^{\epsilon}$ should satisfy the law of energy conservation in Newton's equation,
see Eq. (\ref{x1-x2}), thus yielding
\beq
\label{x1-x2 fixed}
(x_2^{\epsilon})^2 + \omega^2(x_1^{\epsilon})^2+\frac{4}{\pi^2(x_1^{\epsilon})^2}+\frac{4u^{\epsilon}}{\pi^2x_1^{\epsilon}}=c^{\epsilon}.
\eeq

Obviously, the controller $u^{\epsilon}(t)$ (\ref{u_fix}) is a continuous function of $t$, relying on the time-varying $p_2^{\epsilon}$.   Considering the initial and target states, i.e., $(x_1^{\epsilon}(0),x_2^{\epsilon}(0))=(1,0)$, and $(x_1^{\epsilon}(\tau^{\epsilon}),x_2^{\epsilon}(\tau^{\epsilon}))=(\gamma,0)$, we map the controller $u(t)$ (\ref{control sequence u(t)}) into following sequence:
\beq\label{fixed control sequence u(t)}
u^{\epsilon}(t)=%
\begin{cases}
	g_{\mathrm{i}}, \qquad  \qquad \qquad  \qquad   t = 0\\ 
	\frac{(g_{\mathrm{i}}^{\epsilon}-\delta) p_2^{\epsilon}}{2\sqrt{(p_2^{\epsilon}(t))^2+\epsilon^2(p_1^{\epsilon}(t))^2}}, ~~ 0<t<\tau^{\epsilon} \\ 
	g_{\mathrm{f}}, \qquad \qquad  \qquad  \qquad t = \tau^{\epsilon}
\end{cases}.
\eeq%
By substituting this into Eqs. (\ref{x1dotfixed})-(\ref{fixed p2}), we can finally solve the problem with appropriate boundary conditions, see the detailed discussion below. 

\begin{table}[]
	\centering
	\begin{tabular}{lcccc}
		\hline
		$\epsilon$& $g_{\mathrm{i}}^{\epsilon}/g_{\mathrm{i}} $~&$p_2^{\epsilon}(0)$~&$p_2^{\epsilon}(t_1)$~& $c^{\epsilon}(\gamma^{\epsilon},x_2^{\epsilon}(\tau))$\\   
		\hline
		0 ~~& 1 &13.9915~~ &$9.9953\times 10^{-5}$&(2,0)\\
		0.1~~&0.9979 & 14.1224~~ & $7.3087\times 10^{-5}$ &(1.9991,0.0013)\\
		0.2~~& 0.9940& 14.2316~~& $85770\times 10^{-5}$&(1.9995,0.0031)\\
		0.3~~& 0.9896&14.4910~~& $3.2556\times 10^{-5}$&(1.9998,0.0053)\\
		\hline
	\end{tabular}
	\caption{The parameters for shooting method, where we choose $p_1^{\epsilon}(0) = -1$, and other parameter are same as in Fig.~\ref{fig3}.}
	\label{table1}
\end{table}

The central idea of such regulation is the reformulation of ``bang-bang” control by a smooth function in terms of continuous adjoint vector $\textbf{p}_i(t)$. One can see that by introducing $\epsilon$ we smooth out the control function $(\ref{u_fix})$, which drives the interaction $g(t)$ from $\delta$ to $g_{\mathrm{i}}$ at switching times, without sudden change, see Fig. \ref{fig3}(a), where
different $\epsilon$ are applied for producing the smooth regulation. To understand it better, the corresponding trajectories of $(x_1^{\epsilon}, x_2^{\epsilon})$ 
and the adjoint vectors $(p_1^{\epsilon}, p_2^{\epsilon})$ 
are also shown in Fig. \ref{fig3}(b) and (c). In the numerical calculation, we use the continuous controller $u^{\epsilon}(t)$ to solve the coupled differential equations, see Eqs. (\ref{x1dotfixed})-(\ref{fixed p2}) for dynamics and adjoint vector, by using shooting method.  When the controller of ``bang-bang" type is replaced by the regulated one (\ref{u_fix}), the total time $\tau$ and final state are of dependence on the different initial boundary conditions. 
So we have to introduce two assumptions in the numerical calculation. On one hand, the initial boundary conditions for $p_1^{\epsilon}(0)$ and $p_2^{\epsilon}(0)$ should guarantee the maximization of control Hamiltonian $H_c(\textbf{p}^{\epsilon},\textbf{x}^{\epsilon},u^{\epsilon})$, i.e. $p_2^{\epsilon}> 0$ ($p_2^{\epsilon} <0$) when  $t<t_1$ ($t>t_1$). On the other hand, the constant $c^{\epsilon}$ in Eq. (\ref{x1-x2 fixed}) at $t=\tau$, featuring the target state, should be as close as possible to $c(\gamma,0)$. In detail, we take the  $p_1(0)=-1$ and  $p_2(0)=13.9915$ when $\epsilon=0$ as reference. Then we simple fix $p_1^{\epsilon}(0)= -1$ and slightly change $p_2^{\epsilon}(0)$ to fulfill the aforementioned two conditions. By using shooting 
method, we apply the parameters listed in Table \ref{table1}
to achieve the sub-optimal solution with smooth controller, see Fig. \ref{fig3}. 
It turns out that the small deviation $g_{\mathrm{i}}^{\epsilon}$ makes the controller smooth at the cost of operating time $\tau$, with an error of magnitude less than $10^{-3}$, see Table \ref{table1}.

%
	
\section{Discussion}

\begin{figure}[]
	\includegraphics[width=\columnwidth]{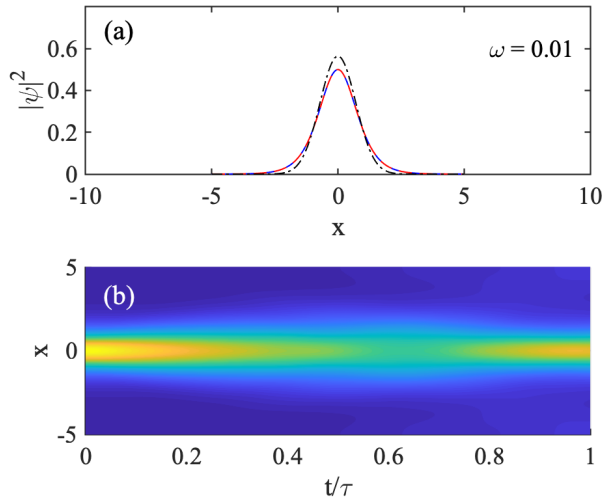}
	\centering
	\caption{(a) Comparison of sech (red dashed) and Gaussian (black dot-dashed) ansatzs with the initial state (blue solid) calculated from imaginary-time method,	where $g_{\mathrm{i}} = -2.0005$, and trap frequency $\omega =0.01$. (b) The state evolution,  $|\psi(x,t)|^2$ numerically calculated from split operator method, is presented with the parameters in ``bang-bang" control, see Fig. \ref{fig1}.}
	\label{fig4}
\end{figure}

In this section, we will perform the numerical calculation. To this aim, the imaginary-time evolution method is used for obtaining the initial and final stationary states, and the state evolving is numerically calculated by means of the split-step method. The validity of sech ansatz (\ref{sech}), comparing with the Gaussian counterpart, is first checked out. In Fig.~\ref{fig4}(a), we confirm that sech ansatz is more accurate than Gaussian one for the problem of soliton compression/decompression, when $\omega \ll 1$. The state evolution, $|\psi(x,t)|^2$, is carried out by using our designed protocols, starting from the initial state, see Fig. \ref{fig4}(b). Remarkably, by using the time-optimal bang-bang control, the bright-soliton matter wave can be expanded within minimal time. However, during the state evolution, the shape of soliton is significantly distorted, resulting from abrupt change of controller $u$, i.e.
the atomic interaction. So the smooth regularization meets the requirement for remedying the difficulties in practical experiments, for instance, the fast adjustment of magnetic field, the induced heating or atom loss following magnetic field ramps across a Feshbach reasonance.

\begin{figure}[]
	\includegraphics[width=\columnwidth]{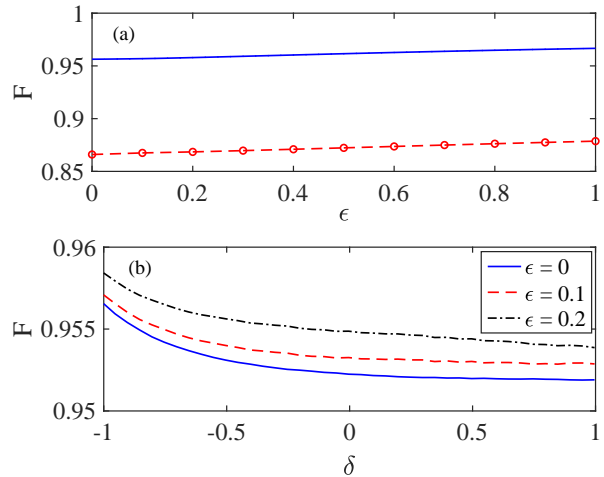}
	\centering
	\caption{(a) Fidelity versus the parameter $\epsilon$ with the protocol designed from smooth regularization.
	Blue solid and red dashed curves present the results obtained from the 1D and 3D simulation, respectively, where the parameters are the same as those in Fig.~\ref{fig1}. (b) Fidelity versus the physical constraint $\delta$, for different $\epsilon$, where $\epsilon = 0$ (blue solid), and $\epsilon = 0.1$ (red dashed), $\epsilon = 0.2$ (black dot-dashed), where other parameters are the same as those in Fig.~\ref{fig1}.}
	\label{fig5}
\end{figure}

To quantify the stability, we define the fidelity as $F=|\langle \psi'_{f}(x)|\psi (x,t_{f})\rangle |^{2}$, where wave function $\psi'_{f}(x)$ is the final stationary state given by the imaginary-time evolution as well. Fig. \ref{fig5}(a) shows that the smooth regulation improves the stability of ``bang-bang" control by smoothing out the controller with the parameter $\epsilon$. Moreover, for larger constrains of $\delta$, the sudden change of atom-atom interaction from negative and positive will make the state evolution unstable. However, the smooth regulation enhances the performance by avoiding the sudden change, see Fig. \ref{fig5}(b), as compared to the case of  ``bang-bang'' control. In other word, one can always shorten the operation time by increasing the constraint $\delta$. But it requires the dramatic change of atom-atom interaction by applying external magnetic field. So, these results demonstrate that there is a trade-off between stability and time, and smooth regulation somehow helps the balance.  

In a realistic BEC experiment, such as quench interaction for
creating bright soliton \cite{khaykovich2002formation} and studying the excitation mode \cite{Excitation2019prl}, we offer an alternative approach for improving unstable experimental conditions. The advantages of smooth ``bang-bang" protocols are two-fold. One one hand, the minimal-time protocol makes the soliton expansion as fast as possible to prevent the atom loss, e.g. from inelastic three-body collisions \cite{collective}. One the other hand, the smooth controller is easy to implement practically, and 
can suppress the heating and atom loss induced from the ramping of interaction. Finally, we emphasize that our model is restricted to an effectively 1D trap with a strong transverse confinement. But one may consider the 
influence of transverse confinement within
the framework of 3D GP equation \cite{Lucapra}, see Fig. \ref{fig5}(a), where the dimensionless $g_{3D}(t)=2\pi g(t)$ in Eq. (\ref{3DGP}) is used in the numerical
calculated, with our designed protocols.

	
\section{Conclusion}
	
In summary, we have studied the variation control of bright soliton matter-wave by manipulating the atomic attraction through Feshbach resonances. By using the variational approximation the motion equation is derived  for capturing the soliton's shape, without dynamical invariant \cite{chenprl104} or Thomas-Fermi limit \cite{muga2009frictionless,StefanatosLiPRA2012,PhysRevR2020}. Sharing with the concept of STA, 
we engineer inversely the atom-atom interaction for achieving the fast but stable soliton decompression within shorter time. 
We apply the Pontryagain's maximum principle in optimal control theory to obtain the minimum-time problem, 
which yields the discontinuous ``bang-bang'' protocol. Furthermore, the smooth regularization 
is further used to smooth out the controller in terms of shooting method. Though we consider quasi-1D soliton expansion as an example, our results presented here can be easily extended to soliton decompression/compression \cite{abdullaev2003adiabatic,li2016shortcut}, by varying either the trap frequency or the interaction strength or both \cite{tangyou,Excitation2019prl}, and other nonlinear optical systems \cite{kong2020shortcuts}, by connecting to other method of enhanced STA working for previously intractable Hamiltonians as well \cite{eSTA}. We find that the experimental relevance can benefit from our smooth time-optimal STA protocols, by suppressing the heating and atom losses.

\begin{acknowledgements}

The work is partially supported from NSFC (12075145, 11474193), SMSTC (2019SHZDZX01-ZX04, 18010500400 and 18ZR1415500), the Program for Eastern Scholar, 
HiQ funding for developing STA (YBN2019115204),
Spanish Government via PGC2018-095113-B-I00 (MCIU/AEI/FEDER, UE), Basque Government via IT986-16, QMiCS (820505), OpenSuperQ (820363) of the EU Flagship on Quantum Technologies, and the EU FET Open Grant Quromorphic (828826). X.C. acknowledges 
the Ram\'{o}n y Cajal program (RYC2017-22482). J.L. acknowledges support from the Okinawa Institute of Science
and Technology Graduate University.

\end{acknowledgements}

\bibliographystyle{apsrev4-1}
\bibliography{ref}

\begin{thebibliography}{53}%
\makeatletter
\providecommand \@ifxundefined [1]{%
 \@ifx{#1\undefined}
}%
\providecommand \@ifnum [1]{%
 \ifnum #1\expandafter \@firstoftwo
 \else \expandafter \@secondoftwo
 \fi
}%
\providecommand \@ifx [1]{%
 \ifx #1\expandafter \@firstoftwo
 \else \expandafter \@secondoftwo
 \fi
}%
\providecommand \natexlab [1]{#1}%
\providecommand \enquote  [1]{``#1''}%
\providecommand \bibnamefont  [1]{#1}%
\providecommand \bibfnamefont [1]{#1}%
\providecommand \citenamefont [1]{#1}%
\providecommand \href@noop [0]{\@secondoftwo}%
\providecommand \href [0]{\begingroup \@sanitize@url \@href}%
\providecommand \@href[1]{\@@startlink{#1}\@@href}%
\providecommand \@@href[1]{\endgroup#1\@@endlink}%
\providecommand \@sanitize@url [0]{\catcode `\\12\catcode `\$12\catcode
  `\&12\catcode `\#12\catcode `\^12\catcode `\_12\catcode `\%12\relax}%
\providecommand \@@startlink[1]{}%
\providecommand \@@endlink[0]{}%
\providecommand \url  [0]{\begingroup\@sanitize@url \@url }%
\providecommand \@url [1]{\endgroup\@href {#1}{\urlprefix }}%
\providecommand \urlprefix  [0]{URL }%
\providecommand \Eprint [0]{\href }%
\providecommand \doibase [0]{http://dx.doi.org/}%
\providecommand \selectlanguage [0]{\@gobble}%
\providecommand \bibinfo  [0]{\@secondoftwo}%
\providecommand \bibfield  [0]{\@secondoftwo}%
\providecommand \translation [1]{[#1]}%
\providecommand \BibitemOpen [0]{}%
\providecommand \bibitemStop [0]{}%
\providecommand \bibitemNoStop [0]{.\EOS\space}%
\providecommand \EOS [0]{\spacefactor3000\relax}%
\providecommand \BibitemShut  [1]{\csname bibitem#1\endcsname}%
\let\auto@bib@innerbib\@empty
\bibitem [{\citenamefont {Anderson}\ \emph {et~al.}(1995)\citenamefont
  {Anderson}, \citenamefont {Ensher}, \citenamefont {Matthews}, \citenamefont
  {Wieman},\ and\ \citenamefont {Cornell}}]{anderson1995observation}%
  \BibitemOpen
  \bibfield  {author} {\bibinfo {author} {\bibfnamefont {M.~H.}\ \bibnamefont
  {Anderson}}, \bibinfo {author} {\bibfnamefont {J.~R.}\ \bibnamefont
  {Ensher}}, \bibinfo {author} {\bibfnamefont {M.~R.}\ \bibnamefont
  {Matthews}}, \bibinfo {author} {\bibfnamefont {C.~E.}\ \bibnamefont
  {Wieman}}, \ and\ \bibinfo {author} {\bibfnamefont {E.~A.}\ \bibnamefont
  {Cornell}},\ }\href {\doibase 10.1126/science.269.5221.198} {\bibfield
  {journal} {\bibinfo  {journal} {Science}\ }\textbf {\bibinfo {volume}
  {269}},\ \bibinfo {pages} {198} (\bibinfo {year} {1995})}\BibitemShut
  {NoStop}%
\bibitem [{\citenamefont {Bradley}\ \emph {et~al.}(1995)\citenamefont
  {Bradley}, \citenamefont {Sackett}, \citenamefont {Tollett},\ and\
  \citenamefont {Hulet}}]{bradley1995evidence}%
  \BibitemOpen
  \bibfield  {author} {\bibinfo {author} {\bibfnamefont {C.~C.}\ \bibnamefont
  {Bradley}}, \bibinfo {author} {\bibfnamefont {C.~A.}\ \bibnamefont
  {Sackett}}, \bibinfo {author} {\bibfnamefont {J.~J.}\ \bibnamefont
  {Tollett}}, \ and\ \bibinfo {author} {\bibfnamefont {R.~G.}\ \bibnamefont
  {Hulet}},\ }\href {\doibase 10.1103/PhysRevLett.75.1687} {\bibfield
  {journal} {\bibinfo  {journal} {Phys. Rev. Lett.}\ }\textbf {\bibinfo
  {volume} {75}},\ \bibinfo {pages} {1687} (\bibinfo {year}
  {1995})}\BibitemShut {NoStop}%
\bibitem [{\citenamefont {Davis}\ \emph {et~al.}(1995)\citenamefont {Davis},
  \citenamefont {Mewes}, \citenamefont {Andrews}, \citenamefont {van Druten},
  \citenamefont {Durfee}, \citenamefont {Kurn},\ and\ \citenamefont
  {Ketterle}}]{davis1995bose}%
  \BibitemOpen
  \bibfield  {author} {\bibinfo {author} {\bibfnamefont {K.~B.}\ \bibnamefont
  {Davis}}, \bibinfo {author} {\bibfnamefont {M.~O.}\ \bibnamefont {Mewes}},
  \bibinfo {author} {\bibfnamefont {M.~R.}\ \bibnamefont {Andrews}}, \bibinfo
  {author} {\bibfnamefont {N.~J.}\ \bibnamefont {van Druten}}, \bibinfo
  {author} {\bibfnamefont {D.~S.}\ \bibnamefont {Durfee}}, \bibinfo {author}
  {\bibfnamefont {D.~M.}\ \bibnamefont {Kurn}}, \ and\ \bibinfo {author}
  {\bibfnamefont {W.}~\bibnamefont {Ketterle}},\ }\href {\doibase
  10.1103/PhysRevLett.75.3969} {\bibfield  {journal} {\bibinfo  {journal}
  {Phys. Rev. Lett.}\ }\textbf {\bibinfo {volume} {75}},\ \bibinfo {pages}
  {3969} (\bibinfo {year} {1995})}\BibitemShut {NoStop}%
\bibitem [{\citenamefont {Dalfovo}\ \emph {et~al.}(1999)\citenamefont
  {Dalfovo}, \citenamefont {Giorgini}, \citenamefont {Pitaevskii},\ and\
  \citenamefont {Stringari}}]{dalfovo1999theory}%
  \BibitemOpen
  \bibfield  {author} {\bibinfo {author} {\bibfnamefont {F.}~\bibnamefont
  {Dalfovo}}, \bibinfo {author} {\bibfnamefont {S.}~\bibnamefont {Giorgini}},
  \bibinfo {author} {\bibfnamefont {L.~P.}\ \bibnamefont {Pitaevskii}}, \ and\
  \bibinfo {author} {\bibfnamefont {S.}~\bibnamefont {Stringari}},\ }\href
  {\doibase 10.1103/RevModPhys.71.463} {\bibfield  {journal} {\bibinfo
  {journal} {Rev. Mod. Phys.}\ }\textbf {\bibinfo {volume} {71}},\ \bibinfo
  {pages} {463} (\bibinfo {year} {1999})}\BibitemShut {NoStop}%
\bibitem [{\citenamefont {Dutton}\ \emph {et~al.}(2001)\citenamefont {Dutton},
  \citenamefont {Budde}, \citenamefont {Slowe},\ and\ \citenamefont
  {Hau}}]{DarkScience2001}%
  \BibitemOpen
  \bibfield  {author} {\bibinfo {author} {\bibfnamefont {Z.}~\bibnamefont
  {Dutton}}, \bibinfo {author} {\bibfnamefont {M.}~\bibnamefont {Budde}},
  \bibinfo {author} {\bibfnamefont {C.}~\bibnamefont {Slowe}}, \ and\ \bibinfo
  {author} {\bibfnamefont {L.~V.}\ \bibnamefont {Hau}},\ }\href {\doibase
  10.1126/science.1062527} {\bibfield  {journal} {\bibinfo  {journal}
  {Science}\ }\textbf {\bibinfo {volume} {293}},\ \bibinfo {pages} {663}
  (\bibinfo {year} {2001})}\BibitemShut {NoStop}%
\bibitem [{\citenamefont {Burger}\ \emph {et~al.}(1999)\citenamefont {Burger},
  \citenamefont {Bongs}, \citenamefont {Dettmer}, \citenamefont {Ertmer},
  \citenamefont {Sengstock}, \citenamefont {Sanpera}, \citenamefont
  {Shlyapnikov},\ and\ \citenamefont {Lewenstein}}]{DarkPRL.83.5198}%
  \BibitemOpen
  \bibfield  {author} {\bibinfo {author} {\bibfnamefont {S.}~\bibnamefont
  {Burger}}, \bibinfo {author} {\bibfnamefont {K.}~\bibnamefont {Bongs}},
  \bibinfo {author} {\bibfnamefont {S.}~\bibnamefont {Dettmer}}, \bibinfo
  {author} {\bibfnamefont {W.}~\bibnamefont {Ertmer}}, \bibinfo {author}
  {\bibfnamefont {K.}~\bibnamefont {Sengstock}}, \bibinfo {author}
  {\bibfnamefont {A.}~\bibnamefont {Sanpera}}, \bibinfo {author} {\bibfnamefont
  {G.~V.}\ \bibnamefont {Shlyapnikov}}, \ and\ \bibinfo {author} {\bibfnamefont
  {M.}~\bibnamefont {Lewenstein}},\ }\href {\doibase
  10.1103/PhysRevLett.83.5198} {\bibfield  {journal} {\bibinfo  {journal}
  {Phys. Rev. Lett.}\ }\textbf {\bibinfo {volume} {83}},\ \bibinfo {pages}
  {5198} (\bibinfo {year} {1999})}\BibitemShut {NoStop}%
\bibitem [{\citenamefont {Khaykovich}\ \emph {et~al.}(2002)\citenamefont
  {Khaykovich}, \citenamefont {Schreck}, \citenamefont {Ferrari}, \citenamefont
  {Bourdel}, \citenamefont {Cubizolles}, \citenamefont {Carr}, \citenamefont
  {Castin},\ and\ \citenamefont {Salomon}}]{khaykovich2002formation}%
  \BibitemOpen
  \bibfield  {author} {\bibinfo {author} {\bibfnamefont {L.}~\bibnamefont
  {Khaykovich}}, \bibinfo {author} {\bibfnamefont {F.}~\bibnamefont {Schreck}},
  \bibinfo {author} {\bibfnamefont {G.}~\bibnamefont {Ferrari}}, \bibinfo
  {author} {\bibfnamefont {T.}~\bibnamefont {Bourdel}}, \bibinfo {author}
  {\bibfnamefont {J.}~\bibnamefont {Cubizolles}}, \bibinfo {author}
  {\bibfnamefont {L.~D.}\ \bibnamefont {Carr}}, \bibinfo {author}
  {\bibfnamefont {Y.}~\bibnamefont {Castin}}, \ and\ \bibinfo {author}
  {\bibfnamefont {C.}~\bibnamefont {Salomon}},\ }\href
  {https://science.sciencemag.org/content/296/5571/1290} {\bibfield  {journal}
  {\bibinfo  {journal} {Science}\ }\textbf {\bibinfo {volume} {296}},\ \bibinfo
  {pages} {1290} (\bibinfo {year} {2002})}\BibitemShut {NoStop}%
\bibitem [{\citenamefont {Strecker}\ \emph {et~al.}(2002)\citenamefont
  {Strecker}, \citenamefont {Partridge}, \citenamefont {Truscott},\ and\
  \citenamefont {Hulet}}]{strecker2002formation}%
  \BibitemOpen
  \bibfield  {author} {\bibinfo {author} {\bibfnamefont {K.~E.}\ \bibnamefont
  {Strecker}}, \bibinfo {author} {\bibfnamefont {G.~B.}\ \bibnamefont
  {Partridge}}, \bibinfo {author} {\bibfnamefont {A.~G.}\ \bibnamefont
  {Truscott}}, \ and\ \bibinfo {author} {\bibfnamefont {R.~G.}\ \bibnamefont
  {Hulet}},\ }\href {\doibase 10.1038/nature747} {\bibfield  {journal}
  {\bibinfo  {journal} {Nature}\ }\textbf {\bibinfo {volume} {417}},\ \bibinfo
  {pages} {150} (\bibinfo {year} {2002})}\BibitemShut {NoStop}%
\bibitem [{\citenamefont {Marchant}\ \emph {et~al.}(2013)\citenamefont
  {Marchant}, \citenamefont {Billam}, \citenamefont {Wiles}, \citenamefont
  {Yu}, \citenamefont {Gardiner},\ and\ \citenamefont
  {Cornish}}]{marchant2013controlled}%
  \BibitemOpen
  \bibfield  {author} {\bibinfo {author} {\bibfnamefont {A.}~\bibnamefont
  {Marchant}}, \bibinfo {author} {\bibfnamefont {T.}~\bibnamefont {Billam}},
  \bibinfo {author} {\bibfnamefont {T.}~\bibnamefont {Wiles}}, \bibinfo
  {author} {\bibfnamefont {M.}~\bibnamefont {Yu}}, \bibinfo {author}
  {\bibfnamefont {S.}~\bibnamefont {Gardiner}}, \ and\ \bibinfo {author}
  {\bibfnamefont {S.}~\bibnamefont {Cornish}},\ }\href {\doibase
  10.1038/ncomms2893} {\bibfield  {journal} {\bibinfo  {journal} {Nature
  Communications}\ }\textbf {\bibinfo {volume} {4}},\ \bibinfo {pages} {1}
  (\bibinfo {year} {2013})}\BibitemShut {NoStop}%
\bibitem [{\citenamefont {Marchant}\ \emph {et~al.}(2016)\citenamefont
  {Marchant}, \citenamefont {Billam}, \citenamefont {Yu}, \citenamefont
  {Rakonjac}, \citenamefont {Helm}, \citenamefont {Polo}, \citenamefont
  {Weiss}, \citenamefont {Gardiner},\ and\ \citenamefont
  {Cornish}}]{PhysRevA.93.021604}%
  \BibitemOpen
  \bibfield  {author} {\bibinfo {author} {\bibfnamefont {A.~L.}\ \bibnamefont
  {Marchant}}, \bibinfo {author} {\bibfnamefont {T.~P.}\ \bibnamefont
  {Billam}}, \bibinfo {author} {\bibfnamefont {M.~M.~H.}\ \bibnamefont {Yu}},
  \bibinfo {author} {\bibfnamefont {A.}~\bibnamefont {Rakonjac}}, \bibinfo
  {author} {\bibfnamefont {J.~L.}\ \bibnamefont {Helm}}, \bibinfo {author}
  {\bibfnamefont {J.}~\bibnamefont {Polo}}, \bibinfo {author} {\bibfnamefont
  {C.}~\bibnamefont {Weiss}}, \bibinfo {author} {\bibfnamefont {S.~A.}\
  \bibnamefont {Gardiner}}, \ and\ \bibinfo {author} {\bibfnamefont {S.~L.}\
  \bibnamefont {Cornish}},\ }\href {\doibase 10.1103/PhysRevA.93.021604}
  {\bibfield  {journal} {\bibinfo  {journal} {Phys. Rev. A}\ }\textbf {\bibinfo
  {volume} {93}},\ \bibinfo {pages} {021604} (\bibinfo {year}
  {2016})}\BibitemShut {NoStop}%
\bibitem [{\citenamefont {Di~Carli}\ \emph
  {et~al.}(2019{\natexlab{a}})\citenamefont {Di~Carli}, \citenamefont
  {Colquhoun}, \citenamefont {Henderson}, \citenamefont {Flannigan},
  \citenamefont {Oppo}, \citenamefont {Daley}, \citenamefont {Kuhr},\ and\
  \citenamefont {Haller}}]{Excitation2019prl}%
  \BibitemOpen
  \bibfield  {author} {\bibinfo {author} {\bibfnamefont {A.}~\bibnamefont
  {Di~Carli}}, \bibinfo {author} {\bibfnamefont {C.~D.}\ \bibnamefont
  {Colquhoun}}, \bibinfo {author} {\bibfnamefont {G.}~\bibnamefont
  {Henderson}}, \bibinfo {author} {\bibfnamefont {S.}~\bibnamefont
  {Flannigan}}, \bibinfo {author} {\bibfnamefont {G.-L.}\ \bibnamefont {Oppo}},
  \bibinfo {author} {\bibfnamefont {A.~J.}\ \bibnamefont {Daley}}, \bibinfo
  {author} {\bibfnamefont {S.}~\bibnamefont {Kuhr}}, \ and\ \bibinfo {author}
  {\bibfnamefont {E.}~\bibnamefont {Haller}},\ }\href {\doibase
  10.1103/PhysRevLett.123.123602} {\bibfield  {journal} {\bibinfo  {journal}
  {Phys. Rev. Lett.}\ }\textbf {\bibinfo {volume} {123}},\ \bibinfo {pages}
  {123602} (\bibinfo {year} {2019}{\natexlab{a}})}\BibitemShut {NoStop}%
\bibitem [{\citenamefont {Martin}\ and\ \citenamefont
  {Ruostekoski}(2012)}]{martin2012quantum}%
  \BibitemOpen
  \bibfield  {author} {\bibinfo {author} {\bibfnamefont {A.}~\bibnamefont
  {Martin}}\ and\ \bibinfo {author} {\bibfnamefont {J.}~\bibnamefont
  {Ruostekoski}},\ }\href {\doibase 10.1088/1367-2630/14/4/043040} {\bibfield
  {journal} {\bibinfo  {journal} {New Journal of Physics}\ }\textbf {\bibinfo
  {volume} {14}},\ \bibinfo {pages} {043040} (\bibinfo {year}
  {2012})}\BibitemShut {NoStop}%
\bibitem [{\citenamefont {Helm}\ \emph {et~al.}(2015)\citenamefont {Helm},
  \citenamefont {Cornish},\ and\ \citenamefont {Gardiner}}]{Sagnac}%
  \BibitemOpen
  \bibfield  {author} {\bibinfo {author} {\bibfnamefont {J.~L.}\ \bibnamefont
  {Helm}}, \bibinfo {author} {\bibfnamefont {S.~L.}\ \bibnamefont {Cornish}}, \
  and\ \bibinfo {author} {\bibfnamefont {S.~A.}\ \bibnamefont {Gardiner}},\
  }\href {\doibase 10.1103/PhysRevLett.114.134101} {\bibfield  {journal}
  {\bibinfo  {journal} {Phys. Rev. Lett.}\ }\textbf {\bibinfo {volume} {114}},\
  \bibinfo {pages} {134101} (\bibinfo {year} {2015})}\BibitemShut {NoStop}%
\bibitem [{\citenamefont {McDonald}\ \emph {et~al.}(2014)\citenamefont
  {McDonald}, \citenamefont {Kuhn}, \citenamefont {Hardman}, \citenamefont
  {Bennetts}, \citenamefont {Everitt}, \citenamefont {Altin}, \citenamefont
  {Debs}, \citenamefont {Close},\ and\ \citenamefont
  {Robins}}]{Interferometerexp}%
  \BibitemOpen
  \bibfield  {author} {\bibinfo {author} {\bibfnamefont {G.~D.}\ \bibnamefont
  {McDonald}}, \bibinfo {author} {\bibfnamefont {C.~C.~N.}\ \bibnamefont
  {Kuhn}}, \bibinfo {author} {\bibfnamefont {K.~S.}\ \bibnamefont {Hardman}},
  \bibinfo {author} {\bibfnamefont {S.}~\bibnamefont {Bennetts}}, \bibinfo
  {author} {\bibfnamefont {P.~J.}\ \bibnamefont {Everitt}}, \bibinfo {author}
  {\bibfnamefont {P.~A.}\ \bibnamefont {Altin}}, \bibinfo {author}
  {\bibfnamefont {J.~E.}\ \bibnamefont {Debs}}, \bibinfo {author}
  {\bibfnamefont {J.~D.}\ \bibnamefont {Close}}, \ and\ \bibinfo {author}
  {\bibfnamefont {N.~P.}\ \bibnamefont {Robins}},\ }\href {\doibase
  10.1103/PhysRevLett.113.013002} {\bibfield  {journal} {\bibinfo  {journal}
  {Phys. Rev. Lett.}\ }\textbf {\bibinfo {volume} {113}},\ \bibinfo {pages}
  {013002} (\bibinfo {year} {2014})}\BibitemShut {NoStop}%
\bibitem [{\citenamefont {Gertjerenken}\ \emph {et~al.}(2013)\citenamefont
  {Gertjerenken}, \citenamefont {Billam}, \citenamefont {Blackley},
  \citenamefont {Le~Sueur}, \citenamefont {Khaykovich}, \citenamefont
  {Cornish},\ and\ \citenamefont {Weiss}}]{Bellstate}%
  \BibitemOpen
  \bibfield  {author} {\bibinfo {author} {\bibfnamefont {B.}~\bibnamefont
  {Gertjerenken}}, \bibinfo {author} {\bibfnamefont {T.~P.}\ \bibnamefont
  {Billam}}, \bibinfo {author} {\bibfnamefont {C.~L.}\ \bibnamefont
  {Blackley}}, \bibinfo {author} {\bibfnamefont {C.~R.}\ \bibnamefont
  {Le~Sueur}}, \bibinfo {author} {\bibfnamefont {L.}~\bibnamefont
  {Khaykovich}}, \bibinfo {author} {\bibfnamefont {S.~L.}\ \bibnamefont
  {Cornish}}, \ and\ \bibinfo {author} {\bibfnamefont {C.}~\bibnamefont
  {Weiss}},\ }\href {\doibase 10.1103/PhysRevLett.111.100406} {\bibfield
  {journal} {\bibinfo  {journal} {Phys. Rev. Lett.}\ }\textbf {\bibinfo
  {volume} {111}},\ \bibinfo {pages} {100406} (\bibinfo {year}
  {2013})}\BibitemShut {NoStop}%
\bibitem [{\citenamefont {Billam}\ \emph {et~al.}(2012)\citenamefont {Billam},
  \citenamefont {Wrathmall},\ and\ \citenamefont {Gardiner}}]{3Danisotropic}%
  \BibitemOpen
  \bibfield  {author} {\bibinfo {author} {\bibfnamefont {T.~P.}\ \bibnamefont
  {Billam}}, \bibinfo {author} {\bibfnamefont {S.~A.}\ \bibnamefont
  {Wrathmall}}, \ and\ \bibinfo {author} {\bibfnamefont {S.~A.}\ \bibnamefont
  {Gardiner}},\ }\href {\doibase 10.1103/PhysRevA.85.013627} {\bibfield
  {journal} {\bibinfo  {journal} {Phys. Rev. A}\ }\textbf {\bibinfo {volume}
  {85}},\ \bibinfo {pages} {013627} (\bibinfo {year} {2012})}\BibitemShut
  {NoStop}%
\bibitem [{\citenamefont {Donley}\ \emph {et~al.}(2001)\citenamefont {Donley},
  \citenamefont {Claussen}, \citenamefont {Cornish}, \citenamefont {Roberts},
  \citenamefont {Cornell},\ and\ \citenamefont {Wieman}}]{donley2001dynamics}%
  \BibitemOpen
  \bibfield  {author} {\bibinfo {author} {\bibfnamefont {E.~A.}\ \bibnamefont
  {Donley}}, \bibinfo {author} {\bibfnamefont {N.~R.}\ \bibnamefont
  {Claussen}}, \bibinfo {author} {\bibfnamefont {S.~L.}\ \bibnamefont
  {Cornish}}, \bibinfo {author} {\bibfnamefont {J.~L.}\ \bibnamefont
  {Roberts}}, \bibinfo {author} {\bibfnamefont {E.~A.}\ \bibnamefont
  {Cornell}}, \ and\ \bibinfo {author} {\bibfnamefont {C.~E.}\ \bibnamefont
  {Wieman}},\ }\href {\doibase 10.1038/35085500} {\bibfield  {journal}
  {\bibinfo  {journal} {Nature}\ }\textbf {\bibinfo {volume} {412}},\ \bibinfo
  {pages} {295} (\bibinfo {year} {2001})}\BibitemShut {NoStop}%
\bibitem [{\citenamefont {Cornish}\ \emph {et~al.}(2006)\citenamefont
  {Cornish}, \citenamefont {Thompson},\ and\ \citenamefont
  {Wieman}}]{PhysRevLett.96.170401}%
  \BibitemOpen
  \bibfield  {author} {\bibinfo {author} {\bibfnamefont {S.~L.}\ \bibnamefont
  {Cornish}}, \bibinfo {author} {\bibfnamefont {S.~T.}\ \bibnamefont
  {Thompson}}, \ and\ \bibinfo {author} {\bibfnamefont {C.~E.}\ \bibnamefont
  {Wieman}},\ }\href {\doibase 10.1103/PhysRevLett.96.170401} {\bibfield
  {journal} {\bibinfo  {journal} {Phys. Rev. Lett.}\ }\textbf {\bibinfo
  {volume} {96}},\ \bibinfo {pages} {170401} (\bibinfo {year}
  {2006})}\BibitemShut {NoStop}%
\bibitem [{\citenamefont {Nguyen}\ \emph {et~al.}(2014)\citenamefont {Nguyen},
  \citenamefont {Dyke}, \citenamefont {Luo}, \citenamefont {Malomed},\ and\
  \citenamefont {Hulet}}]{nguyen2014collisions}%
  \BibitemOpen
  \bibfield  {author} {\bibinfo {author} {\bibfnamefont {J.~H.}\ \bibnamefont
  {Nguyen}}, \bibinfo {author} {\bibfnamefont {P.}~\bibnamefont {Dyke}},
  \bibinfo {author} {\bibfnamefont {D.}~\bibnamefont {Luo}}, \bibinfo {author}
  {\bibfnamefont {B.~A.}\ \bibnamefont {Malomed}}, \ and\ \bibinfo {author}
  {\bibfnamefont {R.~G.}\ \bibnamefont {Hulet}},\ }\href {\doibase
  10.1038/nphys3135} {\bibfield  {journal} {\bibinfo  {journal} {Nature
  Physics}\ }\textbf {\bibinfo {volume} {10}},\ \bibinfo {pages} {918}
  (\bibinfo {year} {2014})}\BibitemShut {NoStop}%
\bibitem [{\citenamefont {Nguyen}\ \emph {et~al.}(2017)\citenamefont {Nguyen},
  \citenamefont {Luo},\ and\ \citenamefont {Hulet}}]{nguyen2017formation}%
  \BibitemOpen
  \bibfield  {author} {\bibinfo {author} {\bibfnamefont {J.~H.}\ \bibnamefont
  {Nguyen}}, \bibinfo {author} {\bibfnamefont {D.}~\bibnamefont {Luo}}, \ and\
  \bibinfo {author} {\bibfnamefont {R.~G.}\ \bibnamefont {Hulet}},\ }\href
  {\doibase 10.1126/science.aal3220} {\bibfield  {journal} {\bibinfo  {journal}
  {Science}\ }\textbf {\bibinfo {volume} {356}},\ \bibinfo {pages} {422}
  (\bibinfo {year} {2017})}\BibitemShut {NoStop}%
\bibitem [{\citenamefont {Di~Carli}\ \emph
  {et~al.}(2019{\natexlab{b}})\citenamefont {Di~Carli}, \citenamefont
  {Colquhoun}, \citenamefont {Henderson}, \citenamefont {Flannigan},
  \citenamefont {Oppo}, \citenamefont {Daley}, \citenamefont {Kuhr},\ and\
  \citenamefont {Haller}}]{breathingmode}%
  \BibitemOpen
  \bibfield  {author} {\bibinfo {author} {\bibfnamefont {A.}~\bibnamefont
  {Di~Carli}}, \bibinfo {author} {\bibfnamefont {C.~D.}\ \bibnamefont
  {Colquhoun}}, \bibinfo {author} {\bibfnamefont {G.}~\bibnamefont
  {Henderson}}, \bibinfo {author} {\bibfnamefont {S.}~\bibnamefont
  {Flannigan}}, \bibinfo {author} {\bibfnamefont {G.-L.}\ \bibnamefont {Oppo}},
  \bibinfo {author} {\bibfnamefont {A.~J.}\ \bibnamefont {Daley}}, \bibinfo
  {author} {\bibfnamefont {S.}~\bibnamefont {Kuhr}}, \ and\ \bibinfo {author}
  {\bibfnamefont {E.}~\bibnamefont {Haller}},\ }\href {\doibase
  10.1103/PhysRevLett.123.123602} {\bibfield  {journal} {\bibinfo  {journal}
  {Phys. Rev. Lett.}\ }\textbf {\bibinfo {volume} {123}},\ \bibinfo {pages}
  {123602} (\bibinfo {year} {2019}{\natexlab{b}})}\BibitemShut {NoStop}%
\bibitem [{\citenamefont {Longenecker}\ and\ \citenamefont
  {Mueller}(2019)}]{collective}%
  \BibitemOpen
  \bibfield  {author} {\bibinfo {author} {\bibfnamefont {D.}~\bibnamefont
  {Longenecker}}\ and\ \bibinfo {author} {\bibfnamefont {E.~J.}\ \bibnamefont
  {Mueller}},\ }\href {\doibase 10.1103/PhysRevA.99.053618} {\bibfield
  {journal} {\bibinfo  {journal} {Phys. Rev. A}\ }\textbf {\bibinfo {volume}
  {99}},\ \bibinfo {pages} {053618} (\bibinfo {year} {2019})}\BibitemShut
  {NoStop}%
\bibitem [{\citenamefont {Torrontegui}\ \emph {et~al.}(2013)\citenamefont
  {Torrontegui}, \citenamefont {Ib{\'a}nez}, \citenamefont
  {Mart{\'\i}nez-Garaot}, \citenamefont {Modugno}, \citenamefont {del Campo},
  \citenamefont {Gu{\'e}ry-Odelin}, \citenamefont {Ruschhaupt}, \citenamefont
  {Chen},\ and\ \citenamefont {Muga}}]{torrontegui2013shortcuts}%
  \BibitemOpen
  \bibfield  {author} {\bibinfo {author} {\bibfnamefont {E.}~\bibnamefont
  {Torrontegui}}, \bibinfo {author} {\bibfnamefont {S.}~\bibnamefont
  {Ib{\'a}nez}}, \bibinfo {author} {\bibfnamefont {S.}~\bibnamefont
  {Mart{\'\i}nez-Garaot}}, \bibinfo {author} {\bibfnamefont {M.}~\bibnamefont
  {Modugno}}, \bibinfo {author} {\bibfnamefont {A.}~\bibnamefont {del Campo}},
  \bibinfo {author} {\bibfnamefont {D.}~\bibnamefont {Gu{\'e}ry-Odelin}},
  \bibinfo {author} {\bibfnamefont {A.}~\bibnamefont {Ruschhaupt}}, \bibinfo
  {author} {\bibfnamefont {X.}~\bibnamefont {Chen}}, \ and\ \bibinfo {author}
  {\bibfnamefont {J.~G.}\ \bibnamefont {Muga}},\ }in\ \href@noop {} {\emph
  {\bibinfo {booktitle} {Advances in atomic, molecular, and optical
  physics}}},\ Vol.~\bibinfo {volume} {62}\ (\bibinfo  {publisher} {Elsevier},\
  \bibinfo {year} {2013})\ pp.\ \bibinfo {pages} {117--169}\BibitemShut
  {NoStop}%
\bibitem [{\citenamefont {Gu\'ery-Odelin}\ \emph {et~al.}(2019)\citenamefont
  {Gu\'ery-Odelin}, \citenamefont {Ruschhaupt}, \citenamefont {Kiely},
  \citenamefont {Torrontegui}, \citenamefont {Mart\'{\i}nez-Garaot},\ and\
  \citenamefont {Muga}}]{STARMP}%
  \BibitemOpen
  \bibfield  {author} {\bibinfo {author} {\bibfnamefont {D.}~\bibnamefont
  {Gu\'ery-Odelin}}, \bibinfo {author} {\bibfnamefont {A.}~\bibnamefont
  {Ruschhaupt}}, \bibinfo {author} {\bibfnamefont {A.}~\bibnamefont {Kiely}},
  \bibinfo {author} {\bibfnamefont {E.}~\bibnamefont {Torrontegui}}, \bibinfo
  {author} {\bibfnamefont {S.}~\bibnamefont {Mart\'{\i}nez-Garaot}}, \ and\
  \bibinfo {author} {\bibfnamefont {J.~G.}\ \bibnamefont {Muga}},\ }\href
  {\doibase 10.1103/RevModPhys.91.045001} {\bibfield  {journal} {\bibinfo
  {journal} {Rev. Mod. Phys.}\ }\textbf {\bibinfo {volume} {91}},\ \bibinfo
  {pages} {045001} (\bibinfo {year} {2019})}\BibitemShut {NoStop}%
\bibitem [{\citenamefont {Edmonds}\ \emph {et~al.}(2018)\citenamefont
  {Edmonds}, \citenamefont {Billam}, \citenamefont {Gardiner},\ and\
  \citenamefont {Busch}}]{Thomas}%
  \BibitemOpen
  \bibfield  {author} {\bibinfo {author} {\bibfnamefont {M.~J.}\ \bibnamefont
  {Edmonds}}, \bibinfo {author} {\bibfnamefont {T.~P.}\ \bibnamefont {Billam}},
  \bibinfo {author} {\bibfnamefont {S.~A.}\ \bibnamefont {Gardiner}}, \ and\
  \bibinfo {author} {\bibfnamefont {T.}~\bibnamefont {Busch}},\ }\href
  {\doibase 10.1103/PhysRevA.98.063626} {\bibfield  {journal} {\bibinfo
  {journal} {Phys. Rev. A}\ }\textbf {\bibinfo {volume} {98}},\ \bibinfo
  {pages} {063626} (\bibinfo {year} {2018})}\BibitemShut {NoStop}%
\bibitem [{\citenamefont {P\'erez-Garc\'{\i}a}\ \emph
  {et~al.}(1996)\citenamefont {P\'erez-Garc\'{\i}a}, \citenamefont {Michinel},
  \citenamefont {Cirac}, \citenamefont {Lewenstein},\ and\ \citenamefont
  {Zoller}}]{variationalprl}%
  \BibitemOpen
  \bibfield  {author} {\bibinfo {author} {\bibfnamefont {V.~M.}\ \bibnamefont
  {P\'erez-Garc\'{\i}a}}, \bibinfo {author} {\bibfnamefont {H.}~\bibnamefont
  {Michinel}}, \bibinfo {author} {\bibfnamefont {J.~I.}\ \bibnamefont {Cirac}},
  \bibinfo {author} {\bibfnamefont {M.}~\bibnamefont {Lewenstein}}, \ and\
  \bibinfo {author} {\bibfnamefont {P.}~\bibnamefont {Zoller}},\ }\href
  {\doibase 10.1103/PhysRevLett.77.5320} {\bibfield  {journal} {\bibinfo
  {journal} {Phys. Rev. Lett.}\ }\textbf {\bibinfo {volume} {77}},\ \bibinfo
  {pages} {5320} (\bibinfo {year} {1996})}\BibitemShut {NoStop}%
\bibitem [{\citenamefont {P\'erez-Garc\'{\i}a}\ \emph
  {et~al.}(1997)\citenamefont {P\'erez-Garc\'{\i}a}, \citenamefont {Michinel},
  \citenamefont {Cirac}, \citenamefont {Lewenstein},\ and\ \citenamefont
  {Zoller}}]{variationalpra}%
  \BibitemOpen
  \bibfield  {author} {\bibinfo {author} {\bibfnamefont {V.~M.}\ \bibnamefont
  {P\'erez-Garc\'{\i}a}}, \bibinfo {author} {\bibfnamefont {H.}~\bibnamefont
  {Michinel}}, \bibinfo {author} {\bibfnamefont {J.~I.}\ \bibnamefont {Cirac}},
  \bibinfo {author} {\bibfnamefont {M.}~\bibnamefont {Lewenstein}}, \ and\
  \bibinfo {author} {\bibfnamefont {P.}~\bibnamefont {Zoller}},\ }\href
  {\doibase 10.1103/PhysRevA.56.1424} {\bibfield  {journal} {\bibinfo
  {journal} {Phys. Rev. A}\ }\textbf {\bibinfo {volume} {56}},\ \bibinfo
  {pages} {1424} (\bibinfo {year} {1997})}\BibitemShut {NoStop}%
\bibitem [{\citenamefont {Li}\ \emph {et~al.}(2016)\citenamefont {Li},
  \citenamefont {Sun},\ and\ \citenamefont {Chen}}]{li2016shortcut}%
  \BibitemOpen
  \bibfield  {author} {\bibinfo {author} {\bibfnamefont {J.}~\bibnamefont
  {Li}}, \bibinfo {author} {\bibfnamefont {K.}~\bibnamefont {Sun}}, \ and\
  \bibinfo {author} {\bibfnamefont {X.}~\bibnamefont {Chen}},\ }\href {\doibase
  10.1038/srep38258} {\bibfield  {journal} {\bibinfo  {journal} {Scientific
  Reports}\ }\textbf {\bibinfo {volume} {6}},\ \bibinfo {pages} {38258}
  (\bibinfo {year} {2016})}\BibitemShut {NoStop}%
\bibitem [{\citenamefont {Li}\ \emph {et~al.}(2018)\citenamefont {Li},
  \citenamefont {Fogarty}, \citenamefont {Campbell}, \citenamefont {Chen},\
  and\ \citenamefont {Busch}}]{jingnpj}%
  \BibitemOpen
  \bibfield  {author} {\bibinfo {author} {\bibfnamefont {J.}~\bibnamefont
  {Li}}, \bibinfo {author} {\bibfnamefont {T.}~\bibnamefont {Fogarty}},
  \bibinfo {author} {\bibfnamefont {S.}~\bibnamefont {Campbell}}, \bibinfo
  {author} {\bibfnamefont {X.}~\bibnamefont {Chen}}, \ and\ \bibinfo {author}
  {\bibfnamefont {{\relax Th}.}~\bibnamefont {Busch}},\ }\href {\doibase
  10.1088/1367-2630/aa9cd8} {\bibfield  {journal} {\bibinfo  {journal} {New J.
  Phys.}\ }\textbf {\bibinfo {volume} {20}},\ \bibinfo {pages} {015005}
  (\bibinfo {year} {2018})}\BibitemShut {NoStop}%
\bibitem [{\citenamefont {Xu}\ \emph {et~al.}(2020)\citenamefont {Xu},
  \citenamefont {Li}, \citenamefont {Busch}, \citenamefont {Chen},\ and\
  \citenamefont {Fogarty}}]{xu2019quantum}%
  \BibitemOpen
  \bibfield  {author} {\bibinfo {author} {\bibfnamefont {T.-N.}\ \bibnamefont
  {Xu}}, \bibinfo {author} {\bibfnamefont {J.}~\bibnamefont {Li}}, \bibinfo
  {author} {\bibfnamefont {T.}~\bibnamefont {Busch}}, \bibinfo {author}
  {\bibfnamefont {X.}~\bibnamefont {Chen}}, \ and\ \bibinfo {author}
  {\bibfnamefont {T.}~\bibnamefont {Fogarty}},\ }\href {\doibase
  10.1103/PhysRevResearch.2.023125} {\bibfield  {journal} {\bibinfo  {journal}
  {Phys. Rev. Research}\ }\textbf {\bibinfo {volume} {2}},\ \bibinfo {pages}
  {023125} (\bibinfo {year} {2020})}\BibitemShut {NoStop}%
\bibitem [{\citenamefont {Huang}\ \emph {et~al.}(2020)\citenamefont {Huang},
  \citenamefont {Malomed},\ and\ \citenamefont {Chen}}]{tangyou}%
  \BibitemOpen
  \bibfield  {author} {\bibinfo {author} {\bibfnamefont {T.-Y.}\ \bibnamefont
  {Huang}}, \bibinfo {author} {\bibfnamefont {B.~A.}\ \bibnamefont {Malomed}},
  \ and\ \bibinfo {author} {\bibfnamefont {X.}~\bibnamefont {Chen}},\ }\href
  {\doibase 10.1063/5.0004309} {\bibfield  {journal} {\bibinfo  {journal}
  {Chaos: An Interdisciplinary Journal of Nonlinear Science}\ }\textbf
  {\bibinfo {volume} {30}},\ \bibinfo {pages} {053131} (\bibinfo {year}
  {2020})}\BibitemShut {NoStop}%
\bibitem [{\citenamefont {Muga}\ \emph {et~al.}(2009)\citenamefont {Muga},
  \citenamefont {Chen}, \citenamefont {Ruschhaupt},\ and\ \citenamefont
  {Gu{\'e}ry-Odelin}}]{muga2009frictionless}%
  \BibitemOpen
  \bibfield  {author} {\bibinfo {author} {\bibfnamefont {J.}~\bibnamefont
  {Muga}}, \bibinfo {author} {\bibfnamefont {X.}~\bibnamefont {Chen}}, \bibinfo
  {author} {\bibfnamefont {A.}~\bibnamefont {Ruschhaupt}}, \ and\ \bibinfo
  {author} {\bibfnamefont {D.}~\bibnamefont {Gu{\'e}ry-Odelin}},\ }\href
  {\doibase 10.1088/0953-4075/42/24/241001} {\bibfield  {journal} {\bibinfo
  {journal} {Journal of Physics B: Atomic, Molecular and Optical Physics}\
  }\textbf {\bibinfo {volume} {42}},\ \bibinfo {pages} {241001} (\bibinfo
  {year} {2009})}\BibitemShut {NoStop}%
\bibitem [{\citenamefont {Chen}\ \emph {et~al.}(2010)\citenamefont {Chen},
  \citenamefont {Ruschhaupt}, \citenamefont {Schmidt}, \citenamefont {del
  Campo}, \citenamefont {Gu\'ery-Odelin},\ and\ \citenamefont
  {Muga}}]{chenprl104}%
  \BibitemOpen
  \bibfield  {author} {\bibinfo {author} {\bibfnamefont {X.}~\bibnamefont
  {Chen}}, \bibinfo {author} {\bibfnamefont {A.}~\bibnamefont {Ruschhaupt}},
  \bibinfo {author} {\bibfnamefont {S.}~\bibnamefont {Schmidt}}, \bibinfo
  {author} {\bibfnamefont {A.}~\bibnamefont {del Campo}}, \bibinfo {author}
  {\bibfnamefont {D.}~\bibnamefont {Gu\'ery-Odelin}}, \ and\ \bibinfo {author}
  {\bibfnamefont {J.~G.}\ \bibnamefont {Muga}},\ }\href {\doibase
  10.1103/PhysRevLett.104.063002} {\bibfield  {journal} {\bibinfo  {journal}
  {Phys. Rev. Lett.}\ }\textbf {\bibinfo {volume} {104}},\ \bibinfo {pages}
  {063002} (\bibinfo {year} {2010})}\BibitemShut {NoStop}%
\bibitem [{\citenamefont {Berry}(2009)}]{berry2009transitionless}%
  \BibitemOpen
  \bibfield  {author} {\bibinfo {author} {\bibfnamefont {M.~V.}\ \bibnamefont
  {Berry}},\ }\href {\doibase 10.1088/1751-8113/42/36/365303} {\bibfield
  {journal} {\bibinfo  {journal} {Journal of Physics A: Mathematical and
  Theoretical}\ }\textbf {\bibinfo {volume} {42}},\ \bibinfo {pages} {365303}
  (\bibinfo {year} {2009})}\BibitemShut {NoStop}%
\bibitem [{\citenamefont {del Campo}(2013)}]{adolfocddriving}%
  \BibitemOpen
  \bibfield  {author} {\bibinfo {author} {\bibfnamefont {A.}~\bibnamefont {del
  Campo}},\ }\href {\doibase 10.1103/PhysRevLett.111.100502} {\bibfield
  {journal} {\bibinfo  {journal} {Phys. Rev. Lett.}\ }\textbf {\bibinfo
  {volume} {111}},\ \bibinfo {pages} {100502} (\bibinfo {year}
  {2013})}\BibitemShut {NoStop}%
\bibitem [{\citenamefont {Deffner}\ \emph {et~al.}(2014)\citenamefont
  {Deffner}, \citenamefont {Jarzynski},\ and\ \citenamefont {del
  Campo}}]{adolfoprx}%
  \BibitemOpen
  \bibfield  {author} {\bibinfo {author} {\bibfnamefont {S.}~\bibnamefont
  {Deffner}}, \bibinfo {author} {\bibfnamefont {C.}~\bibnamefont {Jarzynski}},
  \ and\ \bibinfo {author} {\bibfnamefont {A.}~\bibnamefont {del Campo}},\
  }\href {\doibase 10.1103/PhysRevX.4.021013} {\bibfield  {journal} {\bibinfo
  {journal} {Phys. Rev. X}\ }\textbf {\bibinfo {volume} {4}},\ \bibinfo {pages}
  {021013} (\bibinfo {year} {2014})}\BibitemShut {NoStop}%
\bibitem [{\citenamefont {Masuda}\ and\ \citenamefont
  {Nakamura}(2008)}]{masuda2008fast}%
  \BibitemOpen
  \bibfield  {author} {\bibinfo {author} {\bibfnamefont {S.}~\bibnamefont
  {Masuda}}\ and\ \bibinfo {author} {\bibfnamefont {K.}~\bibnamefont
  {Nakamura}},\ }\href {\doibase 10.1103/PhysRevA.78.062108} {\bibfield
  {journal} {\bibinfo  {journal} {Phys. Rev. A}\ }\textbf {\bibinfo {volume}
  {78}},\ \bibinfo {pages} {062108} (\bibinfo {year} {2008})}\BibitemShut
  {NoStop}%
\bibitem [{\citenamefont {Torrontegui}\ \emph {et~al.}(2012)\citenamefont
  {Torrontegui}, \citenamefont {Mart\'{\i}nez-Garaot}, \citenamefont
  {Ruschhaupt},\ and\ \citenamefont {Muga}}]{fastforward}%
  \BibitemOpen
  \bibfield  {author} {\bibinfo {author} {\bibfnamefont {E.}~\bibnamefont
  {Torrontegui}}, \bibinfo {author} {\bibfnamefont {S.}~\bibnamefont
  {Mart\'{\i}nez-Garaot}}, \bibinfo {author} {\bibfnamefont {A.}~\bibnamefont
  {Ruschhaupt}}, \ and\ \bibinfo {author} {\bibfnamefont {J.~G.}\ \bibnamefont
  {Muga}},\ }\href {\doibase 10.1103/PhysRevA.86.013601} {\bibfield  {journal}
  {\bibinfo  {journal} {Phys. Rev. A}\ }\textbf {\bibinfo {volume} {86}},\
  \bibinfo {pages} {013601} (\bibinfo {year} {2012})}\BibitemShut {NoStop}%
\bibitem [{\citenamefont {Stefanatos}\ and\ \citenamefont
  {Li}(2012)}]{StefanatosLiPRA2012}%
  \BibitemOpen
  \bibfield  {author} {\bibinfo {author} {\bibfnamefont {D.}~\bibnamefont
  {Stefanatos}}\ and\ \bibinfo {author} {\bibfnamefont {J.-S.}\ \bibnamefont
  {Li}},\ }\href {\doibase 10.1103/PhysRevA.86.063602} {\bibfield  {journal}
  {\bibinfo  {journal} {Phys. Rev. A}\ }\textbf {\bibinfo {volume} {86}},\
  \bibinfo {pages} {063602} (\bibinfo {year} {2012})}\BibitemShut {NoStop}%
\bibitem [{\citenamefont {Keller}\ \emph {et~al.}(2020)\citenamefont {Keller},
  \citenamefont {Fogarty}, \citenamefont {Li},\ and\ \citenamefont
  {Busch}}]{PhysRevR2020}%
  \BibitemOpen
  \bibfield  {author} {\bibinfo {author} {\bibfnamefont {T.}~\bibnamefont
  {Keller}}, \bibinfo {author} {\bibfnamefont {T.}~\bibnamefont {Fogarty}},
  \bibinfo {author} {\bibfnamefont {J.}~\bibnamefont {Li}}, \ and\ \bibinfo
  {author} {\bibfnamefont {T.}~\bibnamefont {Busch}},\ }\href {\doibase
  10.1103/PhysRevResearch.2.033335} {\bibfield  {journal} {\bibinfo  {journal}
  {Phys. Rev. Research}\ }\textbf {\bibinfo {volume} {2}},\ \bibinfo {pages}
  {033335} (\bibinfo {year} {2020})}\BibitemShut {NoStop}%
\bibitem [{\citenamefont {Liang}\ \emph {et~al.}(2005)\citenamefont {Liang},
  \citenamefont {Zhang},\ and\ \citenamefont {Liu}}]{LiuPRL}%
  \BibitemOpen
  \bibfield  {author} {\bibinfo {author} {\bibfnamefont {Z.~X.}\ \bibnamefont
  {Liang}}, \bibinfo {author} {\bibfnamefont {Z.~D.}\ \bibnamefont {Zhang}}, \
  and\ \bibinfo {author} {\bibfnamefont {W.~M.}\ \bibnamefont {Liu}},\ }\href
  {\doibase 10.1103/PhysRevLett.94.050402} {\bibfield  {journal} {\bibinfo
  {journal} {Phys. Rev. Lett.}\ }\textbf {\bibinfo {volume} {94}},\ \bibinfo
  {pages} {050402} (\bibinfo {year} {2005})}\BibitemShut {NoStop}%
\bibitem [{\citenamefont {Carr}\ and\ \citenamefont
  {Castin}(2002)}]{CastinPRA}%
  \BibitemOpen
  \bibfield  {author} {\bibinfo {author} {\bibfnamefont {L.~D.}\ \bibnamefont
  {Carr}}\ and\ \bibinfo {author} {\bibfnamefont {Y.}~\bibnamefont {Castin}},\
  }\href {\doibase 10.1103/PhysRevA.66.063602} {\bibfield  {journal} {\bibinfo
  {journal} {Phys. Rev. A}\ }\textbf {\bibinfo {volume} {66}},\ \bibinfo
  {pages} {063602} (\bibinfo {year} {2002})}\BibitemShut {NoStop}%
\bibitem [{\citenamefont {Salasnich}(2004)}]{Lucapra}%
  \BibitemOpen
  \bibfield  {author} {\bibinfo {author} {\bibfnamefont {L.}~\bibnamefont
  {Salasnich}},\ }\href {\doibase 10.1103/PhysRevA.70.053617} {\bibfield
  {journal} {\bibinfo  {journal} {Phys. Rev. A}\ }\textbf {\bibinfo {volume}
  {70}},\ \bibinfo {pages} {053617} (\bibinfo {year} {2004})}\BibitemShut
  {NoStop}%
\bibitem [{\citenamefont {Kirk}(2004)}]{kirk2004optimal}%
  \BibitemOpen
  \bibfield  {author} {\bibinfo {author} {\bibfnamefont {D.~E.}\ \bibnamefont
  {Kirk}},\ }\href@noop {} {\emph {\bibinfo {title} {Optimal control theory: an
  introduction}}}\ (\bibinfo  {publisher} {Courier Corporation},\ \bibinfo
  {year} {2004})\BibitemShut {NoStop}%
\bibitem [{\citenamefont {Ding}\ \emph {et~al.}(2020)\citenamefont {Ding},
  \citenamefont {Huang}, \citenamefont {Paul}, \citenamefont {Hao},\ and\
  \citenamefont {Chen}}]{yongchengPRA}%
  \BibitemOpen
  \bibfield  {author} {\bibinfo {author} {\bibfnamefont {Y.}~\bibnamefont
  {Ding}}, \bibinfo {author} {\bibfnamefont {T.-Y.}\ \bibnamefont {Huang}},
  \bibinfo {author} {\bibfnamefont {K.}~\bibnamefont {Paul}}, \bibinfo {author}
  {\bibfnamefont {M.}~\bibnamefont {Hao}}, \ and\ \bibinfo {author}
  {\bibfnamefont {X.}~\bibnamefont {Chen}},\ }\href {\doibase
  10.1103/PhysRevA.101.063410} {\bibfield  {journal} {\bibinfo  {journal}
  {Phys. Rev. A}\ }\textbf {\bibinfo {volume} {101}},\ \bibinfo {pages}
  {063410} (\bibinfo {year} {2020})}\BibitemShut {NoStop}%
\bibitem [{\citenamefont {Silva}\ and\ \citenamefont
  {Tr{\'e}lat}(2010)}]{silva2010smooth}%
  \BibitemOpen
  \bibfield  {author} {\bibinfo {author} {\bibfnamefont {C.}~\bibnamefont
  {Silva}}\ and\ \bibinfo {author} {\bibfnamefont {E.}~\bibnamefont
  {Tr{\'e}lat}},\ }\href {\doibase 10.1109/TAC.2010.2047742} {\bibfield
  {journal} {\bibinfo  {journal} {IEEE Transactions on Automatic Control}\
  }\textbf {\bibinfo {volume} {55}},\ \bibinfo {pages} {2488} (\bibinfo {year}
  {2010})}\BibitemShut {NoStop}%
\bibitem [{\citenamefont {Landau}\ and\ \citenamefont
  {Lifshitz}(1998)}]{landau1998course}%
  \BibitemOpen
  \bibfield  {author} {\bibinfo {author} {\bibfnamefont {L.}~\bibnamefont
  {Landau}}\ and\ \bibinfo {author} {\bibfnamefont {E.}~\bibnamefont
  {Lifshitz}},\ }\enquote {\bibinfo {title} {Course of theoretical physics.
  vol. 1: Mechanics},}\ \ (\bibinfo  {publisher} {Oxford:
  Butterworth-Heinemann},\ \bibinfo {year} {1998})\BibitemShut {NoStop}%
\bibitem [{\citenamefont {Abdullaev}\ and\ \citenamefont
  {Salerno}(2003)}]{abdullaev2003adiabatic}%
  \BibitemOpen
  \bibfield  {author} {\bibinfo {author} {\bibfnamefont {F.~K.}\ \bibnamefont
  {Abdullaev}}\ and\ \bibinfo {author} {\bibfnamefont {M.}~\bibnamefont
  {Salerno}},\ }\href@noop {} {\bibfield  {journal} {\bibinfo  {journal}
  {Journal of Physics B: Atomic, Molecular and Optical Physics}\ }\textbf
  {\bibinfo {volume} {36}},\ \bibinfo {pages} {2851} (\bibinfo {year}
  {2003})}\BibitemShut {NoStop}%
\bibitem [{\citenamefont {Lu}\ \emph {et~al.}(2014)\citenamefont {Lu},
  \citenamefont {Chen}, \citenamefont {Alonso},\ and\ \citenamefont
  {Muga}}]{LuPRA2014}%
  \BibitemOpen
  \bibfield  {author} {\bibinfo {author} {\bibfnamefont {X.-J.}\ \bibnamefont
  {Lu}}, \bibinfo {author} {\bibfnamefont {X.}~\bibnamefont {Chen}}, \bibinfo
  {author} {\bibfnamefont {J.}~\bibnamefont {Alonso}}, \ and\ \bibinfo {author}
  {\bibfnamefont {J.~G.}\ \bibnamefont {Muga}},\ }\href {\doibase
  10.1103/PhysRevA.89.023627} {\bibfield  {journal} {\bibinfo  {journal} {Phys.
  Rev. A}\ }\textbf {\bibinfo {volume} {89}},\ \bibinfo {pages} {023627}
  (\bibinfo {year} {2014})}\BibitemShut {NoStop}%
\bibitem [{\citenamefont {Stefanatos}\ \emph {et~al.}(2010)\citenamefont
  {Stefanatos}, \citenamefont {Ruths},\ and\ \citenamefont
  {Li}}]{StefanatosPRA2010}%
  \BibitemOpen
  \bibfield  {author} {\bibinfo {author} {\bibfnamefont {D.}~\bibnamefont
  {Stefanatos}}, \bibinfo {author} {\bibfnamefont {J.}~\bibnamefont {Ruths}}, \
  and\ \bibinfo {author} {\bibfnamefont {J.-S.}\ \bibnamefont {Li}},\ }\href
  {\doibase 10.1103/PhysRevA.82.063422} {\bibfield  {journal} {\bibinfo
  {journal} {Phys. Rev. A}\ }\textbf {\bibinfo {volume} {82}},\ \bibinfo
  {pages} {063422} (\bibinfo {year} {2010})}\BibitemShut {NoStop}%
\bibitem [{\citenamefont {Hoffmann}\ \emph {et~al.}(2011)\citenamefont
  {Hoffmann}, \citenamefont {Salamon}, \citenamefont {Rezek},\ and\
  \citenamefont {Kosloff}}]{hoffmann2011time}%
  \BibitemOpen
  \bibfield  {author} {\bibinfo {author} {\bibfnamefont {K.}~\bibnamefont
  {Hoffmann}}, \bibinfo {author} {\bibfnamefont {P.}~\bibnamefont {Salamon}},
  \bibinfo {author} {\bibfnamefont {Y.}~\bibnamefont {Rezek}}, \ and\ \bibinfo
  {author} {\bibfnamefont {R.}~\bibnamefont {Kosloff}},\ }\href {\doibase
  10.1209/0295-5075/96/60015} {\bibfield  {journal} {\bibinfo  {journal} {EPL
  (Europhysics Letters)}\ }\textbf {\bibinfo {volume} {96}},\ \bibinfo {pages}
  {60015} (\bibinfo {year} {2011})}\BibitemShut {NoStop}%
\bibitem [{\citenamefont {Kong}\ \emph {et~al.}(2020)\citenamefont {Kong},
  \citenamefont {Ying},\ and\ \citenamefont {Chen}}]{kong2020shortcuts}%
  \BibitemOpen
  \bibfield  {author} {\bibinfo {author} {\bibfnamefont {Q.}~\bibnamefont
  {Kong}}, \bibinfo {author} {\bibfnamefont {H.}~\bibnamefont {Ying}}, \ and\
  \bibinfo {author} {\bibfnamefont {X.}~\bibnamefont {Chen}},\ }\href@noop {}
  {\bibfield  {journal} {\bibinfo  {journal} {Entropy}\ }\textbf {\bibinfo
  {volume} {22}},\ \bibinfo {pages} {673} (\bibinfo {year} {2020})}\BibitemShut
  {NoStop}%
\bibitem [{\citenamefont {Whitty}\ \emph {et~al.}(2020)\citenamefont {Whitty},
  \citenamefont {Kiely},\ and\ \citenamefont {Ruschhaupt}}]{eSTA}%
  \BibitemOpen
  \bibfield  {author} {\bibinfo {author} {\bibfnamefont {C.}~\bibnamefont
  {Whitty}}, \bibinfo {author} {\bibfnamefont {A.}~\bibnamefont {Kiely}}, \
  and\ \bibinfo {author} {\bibfnamefont {A.}~\bibnamefont {Ruschhaupt}},\
  }\href {\doibase 10.1103/PhysRevResearch.2.023360} {\bibfield  {journal}
  {\bibinfo  {journal} {Phys. Rev. Research}\ }\textbf {\bibinfo {volume}
  {2}},\ \bibinfo {pages} {023360} (\bibinfo {year} {2020})}\BibitemShut
  {NoStop}%
\end{thebibliography}%
\end{document}